\documentclass{appolb}
\usepackage{epsfig}

\begin{document}
\title{Hadronic Effective Field Theory Applied to $\Lambda$-Hypernuclei}
\author{Jeff McIntire
\address{Department of Physics, College of William and Mary,
Williamsburg, VA 23187}
}
\maketitle
\begin{abstract}
In the present work, the approach of Furnstahl, Serot, and Tang (FST) 
is extended to the region of nonzero strangeness in application to 
single-particle states in single $\Lambda$-hypernuclei. To include 
$\Lambda$'s, an additional contribution to their effective lagrangian 
is systematically constructed within the framework of FST. The 
relativistic Hartree (Kohn-Sham) equations are solved numerically, and 
least-square fits to a series of experimental levels are performed
at various levels of truncation in the extended lagrangian. The ground-state
properties of any $\Lambda$-hypernuclei are then predicted. In addition, 
ground-state $\Lambda$-particle--nucleon-hole splittings are 
calculated where appropriate, and the approach is calibrated against
a calculation of the $\mathrm{s}_{1/2}$-doublet splitting
in the nucleus $^{32}_{15}\mathrm{P}_{17}$. 
\end{abstract}
  
\section{Introduction}

Effective field theories have been developed in recent years 
to solve the nuclear many-body problem. In the present work, we consider 
one of these theories, proposed by Furnstahl, Serot, and Tang (FST)
\cite{ref:Fu97,ref:Se97}, and extend their methodology to the 
case of single $\Lambda$-hypernuclei. Specifically, the
phenomena of interest here are ground-state (GS) binding energies,
densities, single-particle spectra, and particle-hole
splittings of select single $\Lambda$-hypernuclei.

FST develop a self-consistent framework for constructing an effective 
lagrangian that incorporates the principles of both quantum mechanics and
special relativity, the underlying symmetries of QCD, and the nonlinear 
realization of spontaneously broken chiral symmetry \cite{ref:Fu97}. 
As this is a low-energy theory, the appropriate low lying hadrons are 
used as degrees of freedom. In order to make any meaningful calculation, 
the lagrangian, which in principle contains an infinite number of terms, must
be truncated in some way. Naive dimensional analysis (NDA) 
\cite{ref:Ge84,ref:Ge93} and relativistic mean field theory (RMFT) 
\cite{ref:Se97,ref:Wa95} provide a formalism in which higher order
terms are, in general, successively smaller; this allows for a systematic 
expansion, and consequently a meaningful truncation, in the 
effective lagrangian. Here FST utilize relativistic Hartree theory to reduce 
the many-body equations to single-particle equations. 
The free parameters in the effective lagrangian are fixed via 
least-squares fits to experimental data from ordinary nuclei
along the valley of stability. These fits are conducted
at various levels of truncation in the underlying lagrangian \cite{ref:Fu97}. 
Once the values of these parameters are known, this lagrangian can be used 
to predict other properties of ordinary nuclei. One example  
which demonstrates the predictive power of
this method is its application to the study of nuclei 
far from stability \cite{ref:He02,ref:He03a}.

Density functional theory (DFT) is a theoretical 
framework which allows one to calculate the GS
properties of many-body systems without carrying around all the baggage
contained in the many-particle wave functions \cite{ref:Ko99}. 
Two points are of interest here. First, if the expectation value 
of the hamiltonian is considered as a functional of the density, the 
exact GS density can be determined by minimizing the energy functional. Second,
one only needs to solve a series of self-consistent, single-particle 
equations with classical fields, instead of many-body
equations with quantum fields \cite{ref:Se01}.
In other words, Kohn-Sham theory is formally equivalent to 
relativistic Hartree theory. Consequently, the problem is now reduced to 
determining the correct form of the energy functional, which
follows from the appropriate lagrangian.
The full interacting lagrangian of FST gives an 
appropriate energy functional and, as a result, DFT 
provides an underlying theoretical justification for this approach.

Hadronic effective lagrangians using MFT have been developed
in the literature to describe hypernuclei. Early models containing 
only the lowest order terms required much
weaker meson couplings to the $\Lambda$ than to the nucleons
to achieve success \cite{ref:Br77,ref:Bo81}, particularly in the 
weak spin-orbit interaction. Later, it was suggested that
large meson couplings to the $\Lambda$ consistent with
$\mathrm{SU(3)}$ were possible if the lagrangian was extended
to include tensor couplings 
\cite{ref:No80,ref:Ma94,ref:Co92,ref:Co91,ref:Je90,ref:Lo95,ref:Gl93}.
It turns out the spin-orbit splitting is very sensitive to the 
size of the tensor coupling to the vector field. The approach of FST
has also been applied to strange hadronic matter \cite{ref:Zh00}. 
More recently, effective theories consistent with 
$\mathrm{SU(3)_{L} \otimes SU(3)_{R}}$ have been constructed
\cite{ref:Pa98,ref:Pa99,ref:Be02}. Another model of interest uses
strangeness changing response functions to calculate 
the spectra of $\mathrm{^{16}_{Y}O}$ and $\mathrm{^{40}_{Y}Ca}$;\footnote{
Here Y denotes a hyperon.} the resulting GS particle-hole 
splittings are small \cite{ref:Mu01}. Other studies include models 
that couple the mesons self-consistently to the quarks within the baryons 
\cite{ref:Sh02,ref:Ts97} and a density dependent relativistic 
hadronic field theory \cite{ref:Ke00}.

The following studies have attempted to fit potentials to the 
hyperon-nucleon interaction. Experimental data has been analyzed 
to obtain a nonlocal and density-dependent $\Lambda$-nucleus 
potential \cite{ref:Ga88,ref:Ya88}. Global optical potentials 
for $\Lambda$ scattering off nuclei were developed \cite{ref:Co94}.
The hypernuclear mass dependence of the binding energies is
reproduced by a $\Lambda$ moving in a Woods-Saxon potential
\cite{ref:Ha96a}. The Nijmegen group has developed Y-N potentials
based on the assumption of $\mathrm{SU(3)}$ symmetry 
\cite{ref:Ri98,ref:Ri99,ref:Ha99}; this fixes the baryon-meson
coupling constants from N-N scattering fits. Similarly, potentials were
constructed by the Julich group assuming $\mathrm{SU(6)}$ symmetry 
\cite{ref:Re94}. Calculations of hypernuclei using these Nijmegen
or Julich potentials include 
\cite{ref:Ku96,ref:Vi98,ref:Ya00,ref:Cu00,ref:Vi01,ref:Fu99}.
Comparable G-matrix calculations with a $\mathrm{SU(6)}$ quark-model
baryon-baryon interaction \cite{ref:Ko00} and
Skyrme-like hyperon-nucleon potentials \cite{ref:La97} have
also been investigated. Other recent approaches include 
using the Fermi hypernetted chain method \cite{ref:Us99,ref:Ar01} and
using a quark model with one boson exchange potentials
\cite{ref:Ok98}.

Many of these studies achieve a good deal of success. However,
the framework of FST is more comprehensive than these approaches 
as it incorporates directly into a hadronic effective field theory
all of the following: special relativity, quantum mechanics,
the underlying symmetry structure of QCD, and the nonlinear 
realization of spontaneously broken chiral symmetry. Furthermore, 
this methodology is theoretically justified by DFT. Once all the 
parameters are fixed, their lagrangian predicts the GS 
properties of any ordinary nucleus. This approach has had
great success \cite{ref:Fu97,ref:He02,ref:He03a}.
Therefore, it is of interest to extend this methodology, 
with all of its intrinsic strengths, to the strangeness 
sector, as is done here. 

In the present work, the approach developed by FST is expanded to the 
region of the strangeness sector that corresponds to $\Lambda$-hypernuclei 
with $\mathrm{S=-1}$ and $\mathrm{T=0}$. To this end, 
we include a single, isoscalar $\Lambda$ field in the theory.\footnote{
The $\Sigma$ is not explicitly included in
the present calculation. An idea of the possible impact of 
$\Lambda - \Sigma$ mixing can be taken from \cite{ref:Do95}.
It should be mentioned that if one views 
the scalar meson as a two-pion resonance, then the $\Sigma$ enters 
implicitly as an intermediate state in our formalism.}
Now, a $\Lambda$-lagrangian is constructed as an additional contribution
to the full interacting effective lagrangian of FST, consistent 
with their methodology. Since the $\Lambda$ is an isoscalar, it does not couple
to either a single Yukawa pion or the rho meson. 
Furthermore, we confine our theory to the 
mesons already included;\footnote{The kaon is not included 
as a degree of freedom in this work. The reason
is that, as with the pion, the kaon has no mean field and does not 
effect the RMFT calculations.} thus, the meson lagrangian,
which in this approach contains the majority of the complexity,
is unaltered. It has been proposed that
a tensor coupling to the vector field be included to reproduce 
the correct experimental spin-orbit splitting of the p-states in 
$\Lambda$-hypernuclei \cite{ref:No80,ref:Ma94}. As it turns out, 
such a term is a natural extension of our lagrangian in 
this framework. Additional higher order terms are also
included to better approximate the exact energy functional.

Following the methodology of FST, our $\Lambda$-lagrangian
contains a number of free parameters. The constants in both
the nucleon and meson sectors are taken from a FST parameter
set corresponding to their full lagrangian. As before, the remaining 
unconstrained parameters are fixed here via least-squares fits to a series 
of experimental data
\cite{ref:Pi91,ref:Be79,ref:Ha96,ref:Ko02,ref:Ta00,ref:Ca74}. 
The 10 pieces of data used include six GS binding energies, three s-p shell
excitations for the $\Lambda$, and the spin-orbit splitting of the 
p-states in $^{13}_{\Lambda}\mathrm{C}$.
The fits are conducted at four different levels of truncation in the 
$\Lambda$-lagrangian. Once these parameters are fixed, 
this lagrangian can be used to predict 
other properties of single $\Lambda$-hypernuclei. 

One other property that is of interest to calculate here is what we 
refer to as $\mathrm{s}_{1/2}$-splittings. These are GS particle-hole 
splittings of select single $\Lambda$-hypernuclei,
such as $^{16}_{\Lambda}\mathrm{O}$, which have a $\Lambda$ in the 
GS and a hole in the last filled nucleon (proton or neutron) shell. 
For these systems, the angular momenta of the $\Lambda$ and the 
nucleon hole couple to form a doublet. 
The size of these splittings is determined by the difference 
of two particle-hole matrix elements \cite{ref:Fe71}. The 
effective particle-hole interaction utilized here follows
directly from the effective theory of the preceding 
discussion. This interaction, to lowest order, is just  
simple scalar and vector meson exchange \cite{ref:Fu85}.\footnote{
The retention of higher diagrams in the effective 
interaction, particularly those including the tensor coupling 
to the $\Lambda$, is left for future work. Also, it is worth noting
that while the kaon makes no contribution at the mean field
level, kaon exchange may play a role in the effective interaction.
Some idea of the relative contribution of kaon exchange can
be obtained from the Nijmegen potentials \cite{ref:Ri99,ref:Ri99a,ref:Ma89}.
An investigation of the effect of kaon exchange on the 
$\mathrm{s}_{1/2}$-splittings in effective field theory
is also left to future work. \label{foot:1}}
A simple Yukawa spatial dependence is obtained when
retardation is neglected in the meson propagators.
With this exception, the full Lorentz structure is maintained \cite{ref:Fu85}.
For the $\Lambda$-N case, there is no isovector component to the
effective interaction or exchange contribution in the two-body
matrix elements. Through angular momentum relations \cite{ref:Ed57}
and some algebra, the matrix elements are reduced to radial 
Slater integrals. Using the Hartree wave functions 
from the $\Lambda$ single-particle calculations to 
evaluate the integrals, these matrix elements, and consequently
the $\mathrm{s_{1/2}}$-splitting, can now be fully determined. 
Once the parameters in the $\Lambda$-lagrangian are known, 
the effective particle-hole interaction is completely 
specified in this approach. In the case of $\mathrm{s_{1/2}}$-splittings in 
$\Lambda$-hypernuclei, only the spatial part of the vector exchange
contributes to the splitting. 
Predictions are given for the GS doublet splittings of every one of the 
$\Lambda$-hypernuclei considered here; all of the doublets used in the 
fitting procedure lie within the current experimental error bars on the 
GS energies. An upcoming high resolution experiment at Jefferson Lab 
will measure the $\mathrm{s_{1/2}}$-splittings in
$^{12}_{\Lambda}\mathrm{B}$ and $^{16}_{\Lambda}\mathrm{N}$ 
\cite{ref:Ga94,ref:Ur01}. The present calculations provide 
theoretical predictions for these quantities.$^{\ref{foot:1}}$
Non-relativistic calculations of 
similar particle-hole splittings have been carried out \cite{ref:Wa71}.

The need for isovector interactions and exchange contributions make 
calculations of similar splittings in ordinary nuclei far more 
complicated \cite{ref:Fu85}. As an example of a comparable system in an 
ordinary nucleus, and to at least partially calibrate the present 
approach, the calculation of the $\mathrm{s_{1/2}}$-splitting in 
$^{32}_{15}\mathrm{P}_{17}$ is included here. Comparable systems for 
ordinary nuclei have also been examined \cite{ref:De61}.

In section \ref{sec:2}, we review the methodology of FST and in section
\ref{sec:3}, we describe the development of our $\Lambda$-lagrangian.
The framework for calculating the particle-hole splittings
is discussed in section \ref{sec:4}. The results of the parameter
fits, single-particle calculations, and $\mathrm{s_{1/2}}$-splittings
are given in section \ref{sec:5}.

\section{Methodology of FST \label{sec:2}}

In this section we review the methodology of FST.
They approach the nuclear many-body problem by 
constructing an effective field theory that retains the 
underlying symmetries of QCD as well as the principles of both special
relativity and quantum mechanics \cite{ref:Fu97}. At 
low-energy, hadrons are the desired degrees of freedom and the ones 
which FST use to construct an effective lagrangian. The nonlinear 
realization of spontaneously broken chiral symmetry is illustrated through
a system of pions, nucleons, and rho mesons. They incorporate
Goldstone pions through the field
\begin{equation}
\mathrm{U(x_{\mu}) \equiv \xi(x_{\mu})\mathbf{1}\xi(x_{\mu})}
= e^{i\pi(\mathrm{x_{\mu}})/f_{\pi}}
\mathbf{1}e^{i\pi(\mathrm{x_{\mu}})/f_{\pi}}
\end{equation}

\noindent where the pion field, $\pi(\mathrm{x_{\mu}})=\frac{1}{2}\vec{\tau}
\cdot \vec{\pi}$, appears to all orders, $\tau$ is a Pauli
matrix, and $f_{\pi}$ is the pion-decay constant \cite{ref:Fu97}. 
An isodoublet nucleon field is included, represented by
\begin{equation} 
\mathrm{N(x_{\mu})} = \left( \begin{array}{c} \mathrm{p(x_{\mu})} 
\\ \mathrm{n(x_{\mu})} 
\end{array} \right)
\end{equation}

\noindent The upper (lower) component corresponds to 
the proton (neutron). To account for the symmetry energy in 
nuclear matter, an isovector-vector rho meson,
$\rho_{\nu}(\mathrm{x_{\mu}}) = \frac{1}{2}\vec{\tau} \cdot \vec{\rho}$, is 
also included.

The following boson fields are also incorporated into this framework,
the first two of which are isoscalar chiral singlets.
A scalar field, $\phi$, is included to simulate the medium-range 
nuclear attraction. Next, they incorporate a vector meson, 
$\mathrm{V_{\mu}}$, to reproduce the short-range nuclear repulsion. 
Lastly, a photon field, $\mathrm{A_{\mu}}$, is added to 
take into consideration the electromagnetic structure of nuclei.

As all possible combinations of the fields,
consistent with this framework, are included, this lagrangian contains 
an infinite number of terms. To conduct any meaningful calculation, 
this lagrangian needs to be truncated at some level. 
FST utilize both NDA and RMFT to accomplish this.
NDA is a framework which identifies all the dimensional 
factors of a given term. Once these dimensional factors, and some 
appropriate counting factors, are extracted from a term, the remaining 
dimensionless constant is of $O(1)$ \cite{ref:Ge84,ref:Ge93}. 
This assumption is known as ``naturalness.'' RMFT states that 
when the baryon density becomes appropriately large,  
the sources and meson fields can be 
replaced by their expectation values; here, the 
expectation values of the meson fields are just their classical fields
\cite{ref:Wa95}. Then we notice
that while the meson mean fields are large, the ratios of these
fields to the chiral symmetry breaking scale, M, 
are small. Furthermore, the size of derivatives is related to 
$k_{\mathrm{F}}$, which is also small compared to M. 
These effects are shown by \cite{ref:Wa95}
\begin{equation}
\mathrm{\frac{\Phi}{M}  \; , \; \frac{W}{M}} \sim \frac{1}{3};
\;\;\;\;\;\;\;\;\;\;\;\;\;
\frac{k_{\mathrm{F}}}{\mathrm{M}} \sim \frac{1}{4}
\end{equation}

\noindent where the scaled meson mean fields are defined as
\begin{equation}
\Phi(\vec{\mathrm{x}}) \equiv \mathrm{g_{S}\phi_{0};} \;\;\;\;\;\;
\mathrm{W}(\vec{\mathrm{x}}) \equiv \mathrm{g_{V}V_{0};} \;\;\;\;\;\;
\mathrm{R}(\vec{\mathrm{x}}) \equiv \mathrm{g_{\rho}b_{0};} \;\;\;\;\;\; 
\mathrm{A}(\vec{\mathrm{x}}) \equiv \mathrm{eA_{0}}
\end{equation}

\noindent The ordering principle developed by FST is
\begin{equation}
\nu = \frac{n}{2} + b + d
\end{equation}

\noindent where for a given term $\nu$ is the order, 
$n$ is the number of fermion fields, $b$ is the number of 
non-Goldstone bosons, and $d$ is the number of derivatives. 
Now a controlled expansion is performed in which higher order 
terms are, in general, progressively smaller.

Using this ordering principle, they construct an 
effective lagrangian in two parts
\cite{ref:Fu97}
\begin{equation}
\mathrm{{\cal{L}}_{FST}(x_{\mu}) = {\cal{L}}_{N}(x_{\mu}) 
+ {\cal{L}}_{M}(x_{\mu})}
\end{equation}

\noindent The fermion part to order $\nu=4$ is given by
\begin{eqnarray}
\mathrm{{\cal{L}}_{N}(x_{\mu})} & = & - \bar{\mathrm{N}}\left\{\gamma_{\mu}
\left[\frac{\partial}{\partial \mathrm{x_{\mu}}} + iv_{\mu}
- i\mathrm{g_{A}}\gamma_{5}a_{\mu} - i\mathrm{g_{V}V_{\mu}} 
- i\mathrm{g}_{\rho}\rho_{\mu} \right.\right. \nonumber \\ & - & \left.\left.
\frac{i}{2}\mathrm{eA_{\mu}}\left(1 + \tau_{3}\right)\right]
+ \mathrm{\left(M - g_{S}\phi\right)}\right\}\mathrm{N}
+ \frac{f_{\rho}\mathrm{g}_{\rho}}{\mathrm{4M}}
\mathrm{\bar{N}\sigma_{\mu\nu}\rho_{\mu\nu}N}
\nonumber \\ & + &
\frac{f\mathrm{_{V}g_{V}}}{\mathrm{4M}}
\mathrm{\bar{N}\sigma_{\mu\nu}V_{\mu\nu}N}
+ \mathrm{\frac{\kappa_{\pi}}{M}\bar{N}}\sigma_{\mu\nu}v_{\mu\nu}
\mathrm{N} + \mathrm{\frac{e}{4M}\bar{N}\lambda\sigma_{\mu\nu}F_{\mu\nu}N}
\nonumber \\ & + &
\frac{i\mathrm{e}}{\mathrm{2M^{2}}}\mathrm{\bar{N}\gamma_{\mu}
\left(\beta_{S}+\beta_{V}\tau_{3}\right)N\frac{\partial}
{\partial x_{\nu}}F_{\mu\nu}}
\label{eqn:FSL}
\end{eqnarray}

\noindent where $\lambda = \frac{1}{2}\lambda_{\mathrm{p}}(1+\tau_{3})
+\frac{1}{2}\lambda_{\mathrm{n}}(1-\tau_{3})$
and $\lambda_{\mathrm{p}}=1.793$ ($\lambda_{\mathrm{n}}=-1.913$) is the 
anomalous magnetic moment of the proton (neutron). Note that
for the purposes of this work, the conventions of \cite{ref:Wa95}
are used. Here we have defined
\begin{equation}
\mathrm{V_{\mu\nu} = \frac{\partial V_{\nu}}{\partial x_{\mu}}
- \frac{\partial V_{\mu}}{\partial x_{\nu}}}
\end{equation}

\noindent $v_{\mu\nu}$, $\rho_{\mu\nu}$, and 
$\mathrm{F}_{\mu\nu}$ are similarly defined for $v_{\mu}$, 
$\rho_{\mu}$, and $\mathrm{A}_{\mu}$ respectively. Notice that 
the pions only couple to the fermions through the 
combinations
\begin{equation}
v_{\mu} = - \frac{i}{2}\left(\xi^{\dagger}\frac{\partial\xi}
{\partial \mathrm{x}_{\mu}} + \xi\frac{\partial\xi^{\dagger}}
{\partial \mathrm{x}_{\mu}}\right) = v_{\mu}^{\dagger}
\end{equation}
\begin{equation}
a_{\mu} = \frac{i}{2}\left(\xi^{\dagger}\frac{\partial\xi}
{\partial \mathrm{x}_{\mu}} - \xi\frac{\partial\xi^{\dagger}}
{\partial \mathrm{x}_{\mu}}
\right) = a_{\mu}^{\dagger}
\end{equation}

\noindent To lowest order, both $v_{\mu}$ and $a_{\mu}$ 
contain derivatives of the pion field; thus soft pions decouple. 
The meson lagrangian to order $\nu=4$ is
\begin{eqnarray}
\mathrm{{\cal{L}}_{M}(x_{\mu})} & = & \mathrm{-\frac{1}{2}\left(1 
+ \alpha_{1}\frac{g_{S}\phi}{M}\right)\left(\frac{\partial \phi}
{\partial x_{\mu}}\right)^{2}}
-\frac{f_{\pi}^{2}}{4}\mathrm{tr\left(\frac{\partial U}
{\partial x_{\mu}}\frac{\partial U^{\dagger}}{\partial x_{\mu}}\right)} 
- \frac{1}{2}\mathrm{tr}\left(\rho_{\mu\nu}\rho_{\mu\nu}\right) \nonumber 
\\ & - & \mathrm{\frac{1}{4}\left(1 + \alpha_{2}\frac{g_{S}\phi}{M}\right)
V_{\mu\nu}V_{\mu\nu}} - \mathrm{g}_{\rho\pi\pi}
\frac{2f_{\pi}^{2}}{\mathrm{m}_{\rho}^{2}}\mathrm{tr}
\left(\rho_{\mu\nu}v_{\mu\nu}\right) \nonumber \\ & + &
\frac{\mathrm{m_{\pi}^{2}}f_{\pi}^{2}}{4}\mathrm{tr
\left(U+U^{\dagger}-2\right)} - \mathrm{\frac{1}{2}\left(1 
+ \eta_{1}\frac{g_{S}\phi}{M}
+ \frac{\eta_{2}}{2}\frac{g_{S}^{2}\phi^{2}}{M^{2}}\right)
m_{V}^{2}V_{\mu}V_{\mu}} \nonumber \\ & + & 
\mathrm{\frac{1}{4!}\zeta_{0}g_{V}^{2}
\left(V_{\mu}V_{\mu}\right)^{2} - \frac{1}{4}F_{\mu\nu}F_{\mu\nu}}
- \mathrm{\left(1 + \eta_{\rho}\frac{g_{S}\phi}{M}\right)m_{\rho}^{2}
tr\left(\rho_{\mu}\rho_{\mu}\right)} \nonumber \\ & - & 
\mathrm{m_{S}^{2}\phi^{2}\left(
\frac{1}{2} + \frac{\kappa_{3}}{3!}\frac{g_{S}\phi}{M} 
+ \frac{\kappa_{4}}{4!}\frac{g_{S}^{2}\phi^{2}}{M^{2}}\right)}
\label{eqn:MSL}
\end{eqnarray} 

\noindent Terms such as $\mathrm{\bar{N}N}\phi^{2}$ are redundant in
this formulation. This stems from the fact that
FST employ meson field redefinitions; since the 
parameters are free, they are also
just redefined. A detailed description of how this lagrangian was
constructed is presented in \cite{ref:Fu97}.

This still constitutes a system of many-body equations with quantum fields.
FST now employ Hartree theory and RMFT to  
reduce the many-body system to a series of single-particle equations
with classical fields. This is equivalent to Kohn-Sham theory in DFT;
{\it therefore, DFT provides the theoretical justification 
for this methodology}. The single-particle hamiltonian takes the form
\cite{ref:Fu97}
\begin{eqnarray}
\mathrm{h}(\vec{\mathrm{x}}) & = & - i\vec{\alpha}\cdotp\vec{\nabla} 
+ \mathrm{W + \frac{1}{2}\tau_{3}R +\frac{1}{2}\left(1 + \tau_{3}\right)A  
+ \mathrm{\beta\left(M - g_{S}\Phi\right)}} - \frac{i}{\mathrm{2M}}
\lambda\beta\vec{\alpha}\cdotp\vec{\nabla} \mathrm{A} \nonumber \\ 
& - & \frac{i}{\mathrm{2M}}\beta\vec{\alpha}\cdotp
\left(f_{\mathrm{V}}\vec{\nabla}\mathrm{W} + f_{\rho}\frac{1}{2}\tau_{3}
\vec{\nabla} \mathrm{R}\right) + 
\mathrm{\frac{1}{2M^{2}}\left(\beta_{S} +\beta_{V}\tau_{3}\right)
\nabla^{2}A}
\label{eqn:SPH}
\end{eqnarray}

\noindent Since the pion has no mean field
in a spherically symmetric system, all of the pion couplings 
drop out. The Hartree wave functions are of the form
\begin{equation}
\psi_{\alpha}(\vec{\mathrm{x}}) = \mathrm{\frac{1}{r}}\left(
\begin{array}{c} i\mathrm{G}_{a}(\mathrm{r})
\Phi_{\kappa \mathrm{m}} \\ - \mathrm{F}_{a}
(\mathrm{r})\Phi_{-\kappa \mathrm{m}}
\end{array}
\right)\zeta_{t}
\label{eqn:Har1}
\end{equation}

\noindent Here $\mathrm{\alpha = \{a,m\} = \{nlsj,m\}}$,
$\zeta_{t}$ is a two component spinor, 
and $t_{a}$ is 1/2 (-1/2) for protons (neutrons).
The $\Phi_{\kappa \mathrm{m}}$ are the spin spherical harmonics. 
Substituting this wave function into the Dirac equation,
\begin{equation}
\mathrm{h}(\vec{\mathrm{x}})\psi_{\alpha}(\vec{\mathrm{x}}) 
= \mathrm{E}_{a}\psi_{\alpha}(\vec{\mathrm{x}})
\end{equation}

\noindent one arrives at the following radial Hartree equations 
\begin{equation}
\left[\frac{\partial}{\partial \mathrm{r}} 
+ \frac{\kappa}{\mathrm{r}}\right]\mathrm{G}_{a}(\mathrm{r})
- \left[\mathrm{E}_{a} - \mathrm{U_{1} + U_{2}}\right]\mathrm{F}_{a}
(\mathrm{r}) - \mathrm{U_{3}G}_{a}(\mathrm{r}) = 0
\end{equation}
\begin{equation}
\left[\frac{\partial}{\partial \mathrm{r}} 
- \frac{\kappa}{\mathrm{r}}\right]\mathrm{F}_{a}(\mathrm{r})
+ \left[\mathrm{E}_{a} - \mathrm{U_{1} - U_{2}}\right]\mathrm{G}_{a}
(\mathrm{r}) + \mathrm{U_{3}F}_{a}(\mathrm{r}) = 0
\end{equation}

\noindent where the single-particle potentials are
\begin{eqnarray}
\mathrm{U_{1}(r)} & = & \mathrm{W(r)} + t_{a}\mathrm{R(r)} 
+\left(t_{a} + \frac{1}{2}\right)\mathrm{A(r)} 
+ \mathrm{\frac{1}{2M^{2}}}\left(\beta_{\mathrm{S}}
+ 2t_{a}\beta_{\mathrm{V}}\right)\nabla^{2}\mathrm{A(r)} \:\:\:\:\:\:\:
\\ \mathrm{U_{2}(r)} & = & \mathrm{M - \Phi(r)}
\\ \mathrm{U_{3}(r)} & = & \frac{1}{\mathrm{2M}}\left[f_{\mathrm{V}}
\mathrm{\frac{\partial W(r)}{\partial r}} 
+ t_{a}f_{\rho}\mathrm{\frac{\partial R(r)}{\partial r}} \right] \nonumber \\
& + &\mathrm{\frac{1}{2M}\frac{\partial A(r)}{\partial r}} \left[
\frac{\left(\mathrm{\lambda_{p}+\lambda_{n}}\right)}{2}
+ t_{a}(\mathrm{\lambda_{p} -\lambda_{n}})\right]
\end{eqnarray}

\noindent The scalar meson equation is determined by minimizing 
the variational derivative of the effective lagrangian with
respect to the scalar meson field. The
other meson equations are constructed in a similar fashion. These
meson equations are \cite{ref:Fu97}
\begin{eqnarray}
- \nabla^{2}\Phi + \mathrm{m_{S}^{2}}\Phi & = & \mathrm{g_{S}^{2}}
\rho_{\mathrm{S}}(\vec{\mathrm{x}})
- \mathrm{\frac{m_{S}^{2}}{M}\Phi^{2}\left(\frac{\kappa_{3}}{2}
+ \frac{\kappa_{4}}{3!}\frac{\Phi}{M}\right)} \nonumber \\ & + &
\mathrm{\frac{g_{S}^{2}}{2M}\left(\eta_{1} + \eta_{2}\frac{\Phi}
{M}\right)\frac{m_{V}^{2}}{g_{V}^{2}}W^{2}} 
+ \mathrm{\frac{\alpha_{1}}{2M}}\left[\left(\vec{\nabla}\Phi\right)^{2}
+ 2\Phi\nabla^{2}\Phi\right] \nonumber \\ & + & 
\mathrm{\frac{\alpha_{2}g_{S}^{2}}{2Mg_{V}^{2}}}
\left(\vec{\nabla}\mathrm{W}\right)^{2}
+ \mathrm{\frac{g_{S}^{2}\eta_{\rho}}{2M}\frac{m_{\rho}^{2}}
{g_{\rho}^{2}}R^{2}}
\\ \mathrm{- \nabla^{2}W + m_{V}^{2}W} & = & \mathrm{g_{V}^{2}}
\left[\mathrm{\rho_{B}}(\vec{\mathrm{x}}) 
+ \frac{f_{\mathrm{V}}}{2\mathrm{M}}
\vec{\nabla}\cdotp\left(\mathrm{\rho_{B}^{T}}(\vec{\mathrm{x}})
\mathrm{\hat{r}}\right)\right] \nonumber \\ & - & 
\mathrm{\left(\eta_{1} + \frac{\eta_{2}}{2}
\frac{\Phi}{M}\right)\frac{\Phi}{M}m_{V}^{2}W} 
- \mathrm{\frac{1}{3!}\zeta_{0}W^{3}} \nonumber \\ & + & 
\mathrm{\frac{\alpha_{2}}{M}}\left(
\vec{\nabla}\Phi\cdotp\vec{\nabla}\mathrm{W} + \Phi\nabla^{2}\mathrm{W}\right) 
- \mathrm{\frac{e^{2}g_{V}}{3g_{\gamma}}}\mathrm{\rho_{chg}}(\vec{\mathrm{x}})
\\ \mathrm{- \nabla^{2}R + m_{\rho}^{2}R} & = & \frac{1}{2}
\mathrm{g}_{\rho}^{2}\left[\rho_{3}(\vec{\mathrm{x}})+ \frac{f_{\rho}}
{2\mathrm{M}}\vec{\nabla}\cdotp
\left(\mathrm{\rho_{3}^{T}}(\vec{\mathrm{x}})\mathrm{\hat{r}}\right)\right] 
- \mathrm{\eta_{\rho}\frac{\Phi}{M}m_{\rho}^{2}R}
\nonumber \\ & - &
\mathrm{\frac{e^2g_{\rho}}{g_{\gamma}}}\mathrm{\rho_{chg}}(\vec{\mathrm{x}})
\\ \mathrm{- \nabla^{2}A} & = & \mathrm{e^{2}}
\mathrm{\rho_{chg}}(\vec{\mathrm{x}})
\end{eqnarray}

\noindent The baryon sources become the 
densities in the meson equations and are given here by
\cite{ref:Fu97}
\begin{eqnarray}
\mathrm{\rho_{S}}(\vec{\mathrm{x}}) & = & \sum_{a}\frac{2j_{a} + 1}
{4\pi \mathrm{r}^{2}}\left(\mathrm{G}_{a}^{2}(\mathrm{r}) 
- \mathrm{F}_{a}^{2}(\mathrm{r})\right) \\
\mathrm{\rho_{B}}(\vec{\mathrm{x}}) & = & \sum_{a}\frac{2j_{a} + 1}
{4\pi \mathrm{r}^{2}}\left(\mathrm{G}_{a}^{2}(\mathrm{r}) 
+ \mathrm{F}_{a}^{2}(\mathrm{r})\right) \\
\mathrm{\rho_{B}^{T}}(\vec{\mathrm{x}}) & = & \sum_{a}\frac{2j_{a} + 1}
{4\pi \mathrm{r}^{2}}2\mathrm{G}_{a}(\mathrm{r})
\mathrm{F}_{a}(\mathrm{r}) \\
\mathrm{\rho_{3}}(\vec{\mathrm{x}}) & = & \sum_{a}\frac{2j_{a} + 1}
{4\pi \mathrm{r}^{2}}\left(2t_{a}\right)\left(
\mathrm{G}_{a}^{2}(\mathrm{r}) 
+ \mathrm{F}_{a}^{2}(\mathrm{r})\right) \\
\mathrm{\rho_{3}^{T}}(\vec{\mathrm{x}}) & = & \sum_{a}\frac{2j_{a} + 1}
{4\pi \mathrm{r}^{2}}\left(2t_{a}\right)2\mathrm{G}_{a}(\mathrm{r})
\mathrm{F}_{a}(\mathrm{r})
\end{eqnarray}

\noindent The charge density is made up of two components
\begin{equation}
\mathrm{\rho_{chg}}(\vec{\mathrm{x}}) = \mathrm{\rho_{d}}(\vec{\mathrm{x}}) 
+ \mathrm{\rho_{m}}(\vec{\mathrm{x}})
\label{eqn:den1}
\end{equation}

\noindent where the first, the direct nucleon charge density, is
\begin{equation}
\mathrm{\rho_{d}}(\vec{\mathrm{x}}) = \mathrm{\rho_{p}}(\vec{\mathrm{x}}) 
+ \frac{1}{2\mathrm{M}}\vec{\nabla}\cdotp
\left(\rho_{a}^{\mathrm{T}}(\vec{\mathrm{x}})\mathrm{\hat{r}}\right) 
+ \mathrm{\frac{1}{2M^{2}}\left[\beta_{S}
\nabla^{2}\rho_{B} + \beta_{V}\nabla^{2}\rho_{3}\right]}
\label{eqn:den2}
\end{equation}

\noindent and the second, the vector meson contribution, is 
\begin{equation}
\mathrm{\rho_{m}}(\vec{\mathrm{x}}) = 
\mathrm{\frac{1}{g_{\gamma}g_{\rho}}\nabla^{2}R
+\frac{1}{3g_{\gamma}g_{V}}\nabla^{2}W}
\end{equation}

\noindent The point proton and nucleon tensor densities in Eq.\
(\ref{eqn:den2}) are
\begin{eqnarray}
\mathrm{\rho_{p}}(\vec{\mathrm{x}}) & = & \frac{1}{2}\sum_{a}\frac{2j_{a} + 1}
{4\pi \mathrm{r}^{2}}(1+2t_{a})(\mathrm{G}_{a}^{2}\mathrm{(r)}+
\mathrm{F}_{a}^{2}(\mathrm{r})) \nonumber \\ 
& = & \frac{1}{2}\left(\rho_{\mathrm{B}} + \rho_{3}\right) \\
\rho_{a}^{\mathrm{T}}(\vec{\mathrm{x}}) & = & \sum_{a}\frac{2j_{a} + 1}
{4\pi \mathrm{r}^{2}}2\lambda\mathrm{G}_{a}\mathrm{(r)}
\mathrm{F}_{a}(\mathrm{r})
\end{eqnarray}

\noindent respectively. Finally, the energy functional 
is given by \cite{ref:Fu97}
\begin{equation}
\mathrm{E} = \sum_{a} \mathrm{E}_{a} - \mathrm{\int d^{3}xU_{m}}
\end{equation}

\noindent where
\begin{eqnarray}
\mathrm{U_{m}} & \equiv & -\mathrm{\frac{1}{2}\Phi\rho_{S} +\frac{1}{2}
W}\left(\rho_{\mathrm{B}} +\frac{f_{\mathrm{V}}}{2\mathrm{M}}
\vec{\nabla}\cdotp\mathrm{\rho_{B}^{T}\hat{r}}\right) 
+ \frac{1}{4}\mathrm{R}\left(\rho_{3} + \frac{f_{\rho}}{2\mathrm{M}}
\vec{\nabla}\cdotp\mathrm{\rho_{3}^{T}\hat{r}}\right) + \mathrm{
\frac{1}{2}A\rho_{d}} \nonumber \\ & + & 
\mathrm{\frac{m_{S}^{2}}{g_{S}^{2}}\frac{\Phi^{3}}
{M}\left(\frac{\kappa_{3}}{12} + \frac{\kappa_{4}}{24}\frac{\Phi}{M}
\right) - \frac{\eta_{\rho}}{4}\frac{\Phi}{M}\frac{m_{\rho}^{2}}
{g_{\rho}^{2}}R^{2}} - \mathrm{
\frac{\Phi}{4M}\left(\eta_{1} + \eta_{2}\frac{\Phi}{M}\right)
\frac{m_{V}^{2}}{g_{V}^{2}}W^{2}}
\nonumber \\ & - & \mathrm{\frac{1}{4!g_{V}^{2}}\zeta_{0}W^{4}
+ \frac{\alpha_{1}}{4g_{S}^{2}}\frac{\Phi}{M}\left(\nabla
\Phi\right)^{2} - \frac{\alpha_{2}}{4g_{V}^{2}}\frac{\Phi}{M}
\left(\nabla W\right)^{2}}
\end{eqnarray}

\noindent The radial Hartree equations and the meson 
equations form a system which is solved self-consistently until a 
global convergence is achieved. FST wrote a program to numerically 
solve the coupled, local, nonlinear, differential equations. Huertas 
has written an independent program which reproduces the results of FST 
\cite{ref:He02,ref:He03a}. The free parameters in this system are 
listed in Table \ref{tab:PAR}. These are fit by FST 
to a series of experimental data along the valley of stability
at various levels of truncation in the underlying effective 
lagrangian \cite{ref:Fu97}. The last three parameters are
fit to the electromagnetic properties of the nucleon.
The remaining constants are determined by minimizing 
a least-squares $\chi^{2}$ fit where 29 pieces of experimental data 
were used. The result of a parameter fit corresponding to their full
lagrangian is shown in Table \ref{tab:PAR}. Note that these parameters
do indeed satisfy the naturalness assumption made earlier 
and as a result, higher order terms are successively
smaller. Also, we mention that increasing the level of truncation
beyond that of the G2 parameter set does not significantly 
improve the fit \cite{ref:Fu97}. Once the free parameters are determined,
{\it this lagrangian can be used to predict other properties
of ordinary nuclei} \cite{ref:Fu97,ref:He02,ref:He03a}.

\begin{table}
\begin{center}
\begin{tabular}{|c|c|c|c|c|c|c|} \hline\hline
              &$\mathrm{m_{S}/M}$   &$\mathrm{g_{S}/4\pi}$ 
&$\mathrm{g_{V}/4\pi}$ &$\mathrm{g_{\rho}/4\pi}$ &$\eta_{1}$ 
&$\kappa_{3}$ \\ \hline
$\mathrm{G2}$ &0.55410              &0.83522
&1.01560               &0.75467                  &0.64992 
&3.2467       \\ \hline\hline
              &$\eta_{\rho}$        &$f_{\mathrm{V}}/4$ 
&$f_{\rho}/4$          &$\eta_{2}$               &$\kappa_{4}$ 
&$\zeta_{0}$  \\ \hline
$\mathrm{G2}$ &0.3901               &0.1734 
&0.9619                &0.10975                  &0.63152
&2.6416       \\ \hline\hline 
              &$\mathrm{\beta_{S}}$ &$\mathrm{\beta_{V}}$
&$\alpha_{1}$          &$\alpha_{2}$             & 
&             \\ \hline
$\mathrm{G2}$ &-0.09328             &-0.45964 
&1.7234                &-1.5798                  & 
&             \\ \hline
\end{tabular}
\caption{The G2 parameter set developed by FST \cite{ref:Fu97}.
The first 4 parameters correspond to $\nu =2$, the next 5 to $\nu =3$,
the following 5 to $\nu =4$, and the last 2 to $\nu =5$.}
\label{tab:PAR}
\end{center}
\end{table}

\section{Application to $\mathbf{\Lambda}$-hypernuclei \label{sec:3}}

We now consider an extension of this approach to the 
strangeness sector. The specific phenomena that we seek to investigate
here are GS binding energies (i.e. chemical potentials), 
densities, single-particle spectra, and particle-hole states of single
$\Lambda$-hypernuclei. To this end we add a single, isoscalar
$\Lambda$ to the theory. Note that the $\Lambda$ is also
a chiral singlet. Then, we construct our effective $\Lambda$-lagrangian
as an additional contribution to the full $\nu=4$ lagrangian of FST, 
utilizing their methodology. This lagrangian is
of the form 
\begin{equation}
\mathrm{{\cal{L}}(x_{\mu}) = {\cal{L}}_{FST}(x_{\mu}) 
+ {\cal{L}}_{\Lambda}(x_{\mu})}
\label{eqn:LAM1}
\end{equation}

\noindent Here we restrict ourselves to the mesons 
already incorporated into the theory by FST; therefore, the 
$\Lambda$-lagrangian is confined to the fermion sector. First, we 
consider all possible contributions up to order $\nu = 2$, consistent 
with this approach. Our effective $\Lambda$-lagrangian now takes the form
\begin{equation}
{\cal{L}}_{\Lambda}^{(2)} = - \bar{\Lambda}\left[\gamma_{\mu}\left(
\frac{\partial}{\partial \mathrm{x}_{\mu}} 
- i\mathrm{g_{V\Lambda}V_{\mu}}\right) + \left(\mathrm{M_{\Lambda} 
- g_{S\Lambda}\phi}\right)\right]\Lambda
\end{equation}

\noindent Notice that the coupling constants, $\mathrm{g_{S\Lambda}}$
and $\mathrm{g_{V\Lambda}}$, are free parameters and are different from 
those used in the nucleon case. Single Yukawa rho 
and pion couplings to the $\Lambda$ 
are absent as they do not conserve isospin. Also, 
no electromagnetic coupling is retained to this order as 
$\mathrm{Q}=0$ for the $\Lambda$. Four fermion terms are discussed
in appendix \ref{sec:app}.

However, this lagrangian, to order $\nu = 2$, fails to 
reproduce the small experimental spin-orbit splitting of the
p-states, as in $\mathrm{^{13}_{\Lambda}C}$ \cite{ref:Ko02}. It
was proposed in the literature that tensor couplings of order
$\nu = 3$ be introduced
to correct for this limitation \cite{ref:No80,ref:Ma94}. We add 
tensor couplings to the vector and photon fields, shown by 
\begin{equation}
\mathrm{{\cal{L}}_{\Lambda}^{(T)} = \frac{g_{T\Lambda}g_{V}}
{4M}\bar{\Lambda}\sigma_{\mu\nu}
V_{\mu\nu}\Lambda + \frac{e}{4M}\bar{\Lambda}
\lambda_{\Lambda}\sigma_{\mu\nu}F_{\mu\nu}\Lambda}
\end{equation}

\noindent The constant $\mathrm{g_{T\Lambda}}$ is a free parameter. Here 
$\lambda_{\Lambda} = -0.613$ is the anomalous magnetic moment
of the $\Lambda$. Since we want to make a full expansion 
in our $\Lambda$-lagrangian to order $\nu=3$, consistent with this approach, 
we must also include three additional terms, shown
by the following
\begin{equation}
\mathrm{{\cal{L}}_{\Lambda}^{(N)}} = \mathrm{\mu_{1}\frac{g_{S}^{2}}
{2M}\bar{\Lambda}\Lambda\phi^{2} + \mu_{2}\frac{g_{V}^{2}}
{2M}\bar{\Lambda}\Lambda V_{\mu}V_{\mu}} 
+ i\mathrm{\mu_{3}\frac{g_{S}g_{V}}{M}\bar{\Lambda}
\gamma_{\mu}\Lambda\phi V_{\mu}} 
\end{equation}

\noindent where $\mu_{1}$, $\mu_{2}$, and $\mu_{3}$ 
are three more free parameters.
In the nucleon case, the terms comparable to
these last three were regrouped through
redefinition of the meson fields. However, in the $\Lambda$ case
this is no longer possible unless additional mesons
are added to the theory. A more complete description of how the
terms in the $\Lambda$-lagrangian are chosen is contained in 
appendix \ref{sec:app}. Now our $\Lambda$-lagrangian, complete to order
$\nu = 3$, is
\begin{equation}
{\cal{L}}_{\Lambda} = {\cal{L}}_{\Lambda}^{(2)} 
+ {\cal{L}}_{\Lambda}^{(\mathrm{T})}
+ {\cal{L}}_{\Lambda}^{(\mathrm{N})}
\end{equation}

\noindent Note that our lagrangian in Eq.\ (\ref{eqn:LAM1})
includes all possible terms up to $\nu = 4$ in the 
nucleon and meson sectors as well.\footnote{It is 
of potential interest to consider coupling additional scalar and vector 
mesons, such as the $\mathrm{f_{0}}$ and the $\Phi$, to the 
strangeness density and conserved strangeness 
current respectively. This allows one to eliminate the terms in 
${\cal{L}}_{\Lambda}^{(\mathrm{N})}$ using the equations of motion and
redefinitions of the new fields. However,
the number of additional terms, and their accompanying
free parameters, introduced to $\nu = 3$
make this approach more complex than the present framework. 
Fortunately, the point is relatively unimportant 
for the single $\Lambda$-hypernuclei considered here 
as these new mesons are self-fields of the $\Lambda$.  
If they are included, they would appear only in the 
energy functional and have no effect on the energy eigenvalues;
as the last eigenvalue in this approach is equivalent to the 
total binding energy per baryon for the GS, 
they have no effect on the cases of interest here.}

In the Hartree formalism, we add a new wave function for each new
baryon, given here for the $\Lambda$ by
\begin{equation}
\psi_{\Lambda}(\vec{\mathrm{x}}) = \mathrm{\frac{1}{r}}\left(
\begin{array}{c} i\mathrm{G_{\Lambda}(r)
\Phi_{\kappa m}} \\ - \mathrm{F_{\Lambda}(r)\Phi_{-\kappa m}}
\end{array}
\right)
\end{equation}

\noindent Plugging this wave function into the Dirac equation yields 
the following new pair of Hartree equations
\begin{equation}
\mathrm{\left[\frac{\partial}{\partial r} 
+ \frac{\kappa}{\mathrm{r}}\right]G_{\Lambda}(r)
- \left[E_{\Lambda} - U_{4} + U_{5}\right]F_{\Lambda}(r) - U_{6}
G_{\Lambda}(r) = 0}
\end{equation}
\begin{equation}
\mathrm{\left[\frac{\partial}{\partial r} 
- \frac{\kappa}{\mathrm{r}}\right]F_{\Lambda}(r)
+ \left[E_{\Lambda} - U_{4} - U_{5}\right]G_{\Lambda}(r) + U_{6}
F_{\Lambda}(r) = 0}
\end{equation}

\noindent where the $\Lambda$ single-particle potentials are
\begin{eqnarray}
\mathrm{U}_{4} & = & \mathrm{\frac{g_{V\Lambda}}{g_{V}}W 
- \frac{\mu_{3}}{M}\Phi W} \\ 
\mathrm{U}_{5} & = & \mathrm{M_{\Lambda} - \frac{g_{S\Lambda}}{g_{S}}\Phi 
+ \frac{\mu_{1}}{2M}\Phi^{2} - \frac{\mu_{2}}{2M}W^{2}} \\
\mathrm{U}_{6} & = & \mathrm{\frac{g_{T\Lambda}}{2M}
\frac{\partial W}{\partial r} +
\frac{\lambda_{\Lambda}}{2M}\frac{\partial A}{\partial r}}
\end{eqnarray}

\noindent Since all our additional terms are in the fermion lagrangian, 
the only change to the meson equations are added contributions
to the source terms. The new contributions to the source terms 
arising from the $\Lambda$-lagrangian are
\begin{eqnarray}
\delta\rho\mathrm{_{S}} & = & \mathrm{\frac{1}
{4\pi r^{2}}\left(G_{\Lambda}^{2}(r)- F_{\Lambda}^{2}(r)\right)
\left(\frac{g_{S\Lambda}}{g_{S}} + \frac{\mu_{1}}{M}\Phi\right)} 
\nonumber \\ & - & \mathrm{\frac{1}{4\pi r^{2}}\left(G_{\Lambda}^{2}(r)
+ F_{\Lambda}^{2}(r)\right)\frac{\mu_{3}}{M} W} \\
\delta\rho\mathrm{_{B}} & = & \mathrm{\frac{1}
{4\pi r^{2}}\left(G_{\Lambda}^{2}(r)+ F_{\Lambda}^{2}(r)\right)
\left(\frac{g_{V\Lambda}}{g_{V}} - \frac{\mu_{3}}{M}\Phi\right)} 
\nonumber \\ & - & 
\mathrm{\frac{1}{4\pi r^{2}}\left(G_{\Lambda}^{2}(r)
- F_{\Lambda}^{2}(r)\right)\frac{\mu_{2}}{M} W} \\
\delta\rho\mathrm{_{B}^{T}} & = & \mathrm{\frac{1}
{4\pi r^{2}}2G_{\Lambda}(r)F_{\Lambda}(r)}
\frac{\mathrm{g_{T\Lambda}}}{f_{\mathrm{V}}} \\
\delta\rho_{a}^{\mathrm{T}} & = & \frac{1}{4\pi \mathrm{r}^{2}}
2\lambda_{\Lambda}\mathrm{G}_{\Lambda}\mathrm{(r)}
\mathrm{F}_{\Lambda}(\mathrm{r})
\end{eqnarray}

\noindent The new energy functional is identical in form 
to the one used by FST, with only one additional energy
eigenvalue, $\mathrm{E_{\Lambda}}$. The numerical solution
to the extended set of coupled, local, nonlinear, differential equations
was obtained by extension of a program developed by Huertas
\cite{ref:He02,ref:He03a}. Here we use the parameter sets
of FST for the nucleon and meson sectors. There are six new 
parameters in our $\Lambda$-lagrangian: $\mathrm{g_{S\Lambda}}$, 
$\mathrm{g_{V\Lambda}}$, $\mathrm{g_{T\Lambda}}$, $\mu_{1}$, $\mu_{2}$,
and $\mu_{3}$. Least-squares fits to a series of experimentally
known $\Lambda$ single-particle levels are conducted at various levels 
of truncation in our $\Lambda$-lagrangian, while maintaining the full 
lagrangian of FST to order $\nu = 4$. Now this lagrangian can be used to 
{\it predict} other properties of single $\Lambda$-hypernuclei. One 
application we investigate in the next section is 
$\mathrm{s}_{1/2}$-splittings.

\section{$\mathbf{s_{1/2}}$-doublets \label{sec:4}}

Consider nuclei like $_{\Lambda}^{16}\mathrm{O}$;
the GSs of such systems are particle-hole states. 
One process by which nuclei of this type 
are created is the reaction $\left(\pi^{+},K^{+}\right)$ 
on target nuclei with closed proton and neutron shells
\cite{ref:Pi91,ref:Be79,ref:Ha96}. During the course
of this reaction a neutron is converted into a $\Lambda$. As a result,
a neutron hole is also created which, for the GS, 
inhabits the outermost neutron shell. The angular momentum of the 
$\Lambda$ and the neutron hole couple to form a multiplet. Since
the $\Lambda$ occupies the $1\mathrm{s}_{1/2}$ shell in the GS, 
there are only two states in these multiplets. It is these 
configurations that we refer to as $\mathrm{s}_{1/2}$-doublets. 
The reaction $\mathrm{(e,e'K^{+})}$ is another process
used to create nuclei of this type \cite{ref:Ga94,ref:Ur01}. This process
differs in that a proton hole is
created here and that greater resolution is possible.

In order to calculate the splitting of these doublets, we 
first consider Dirac two-body matrix elements of the forms \cite{ref:Fu85}
\begin{equation}
\langle (\mathrm{n_{1}}l_{1}\mathrm{j_{1}})(\mathrm{n_{2}}l_{2}\mathrm{j_{2}})
\mathrm{JM|V(r_{12})}|(\mathrm{n_{3}}l_{3}\mathrm{j_{3}})
(\mathrm{n_{4}}l_{4}\mathrm{j_{4}})\mathrm{J'M'}\rangle
\label{eqn:A1}
\end{equation}

\noindent and
\begin{equation}
\langle (\mathrm{n_{1}}l_{1}\mathrm{j_{1}})(\mathrm{n_{2}}l_{2}\mathrm{j_{2}})
\mathrm{JM|V(r_{12})}\vec{\sigma}^{(1)}\cdot\vec{\sigma}^{(2)}
|(\mathrm{n_{3}}l_{3}\mathrm{j_{3}})
(\mathrm{n_{4}}l_{4}\mathrm{j_{4}})\mathrm{J'M'}\rangle
\label{eqn:A2}
\end{equation}

\noindent where the single-particle wave functions are specified by 
$\{\mathrm{n}l\mathrm{j}\}$, corresponding to either the upper or 
lower components in Eq.\ (\ref{eqn:Har1}), and $\mathrm{V(r_{12})}$ is 
some effective interaction. Next, we expand this effective interaction
in terms of Legendre polynomials \cite{ref:Fu85}
\begin{eqnarray}
\mathrm{V(r_{12})} & = &
\sum_{k=0}^{\infty}f_{k}(\mathrm{r_{1},r_{2}})
\mathrm{P}_{k}(\cos\theta_{12}) \label{eqn:A8}
\\ & = & \sum_{k=0}^{\infty}f_{k}(\mathrm{r_{1},r_{2}})
\mathrm{C}_{k}(1)\cdot\mathrm{C}_{k}(2)
\end{eqnarray}

\noindent where \cite{ref:Ed57} 
\begin{equation}
\mathrm{C}_{kq} = \left(\frac{4\pi}{2k+1}\right)^{1/2}\mathrm{Y}_{kq}
(\theta,\phi) 
\end{equation}

\noindent Inverting Eq.\ (\ref{eqn:A8}) yields the expression
\begin{equation}
f_{k}(\mathrm{r_{1},r_{2}})=\frac{2k+1}{2}\int_{-1}^{1}
\mathrm{d}(\cos\theta_{12})\mathrm{P}_{k}(\cos\theta_{12})
\mathrm{V(r_{12})}
\label{eqn:A9}
\end{equation}

\noindent In the case of Eq.\ (\ref{eqn:A2}), the effective interaction 
is coupled to Pauli matrices. Therefore, Eq.\ (\ref{eqn:A8}) is 
modified to
\begin{equation}
\mathrm{V(r_{12})}\vec{\sigma}^{(1)}\cdot\vec{\sigma}^{(2)} =
\sum_{k\lambda}(-1)^{k+1-\lambda}f_{k}(\mathrm{r_{1},r_{2}})
\chi_{\lambda}^{(k,1)}(1)\cdot\chi_{\lambda}^{(k,1)}(2)
\end{equation}

\noindent Here $\chi_{\lambda\mu}^{(k,1)}$ are 
$\mathrm{C}_{kq}$ coupled to Pauli matrices, shown by
\begin{equation}
\chi_{\lambda\mu}^{(k,1)}=\sum_{qq'}\mathrm{C}_{kq}\sigma_{1q'}
\langle kq1q'|k1\lambda\mu\rangle
\end{equation}

Now we introduce a specific type of effective interaction.
The form we use here follows directly from the effective lagrangian in the
preceding section and to lowest order, corresponds to
simple Yukawa couplings of both the scalar and vector fields,
given by
\begin{equation}
\mathrm{V(r_{12})} = \gamma_{4}^{(1)}\gamma_{4}^{(2)}\left[
\mathrm{\frac{-g_{S}g_{S\Lambda}}{4\pi}\frac{e^{-m_{S}r_{12}}}{r_{12}}}
+ \mathrm{\gamma_{\mu}^{(1)}\gamma_{\mu}^{(2)}
\frac{g_{V}g_{V\Lambda}}{4\pi}\frac{e^{-m_{V}r_{12}}}{r_{12}}} \right]
\end{equation}

\noindent Here $\mathrm{r_{12}=|\vec{r}_{1}-\vec{r}_{2}|}$. 
This simplistic spatial dependence is possible because
retardation in the meson propagators is neglected, or
$\mathrm{p_{\mu} = (\vec{p},p_{4}) \rightarrow (\vec{p},0)}$.
Otherwise the full Lorentz structure is maintained \cite{ref:Fu85}.
Couplings to the rho and pion fields are absent as
$\mathrm{T}=0$ for the $\Lambda$. In this formalism, we
can now write
\begin{equation}
f_{k}(\mathrm{r_{1},r_{2}}) = \gamma_{4}^{(1)}\gamma_{4}^{(2)}
\left[f_{k}^{\mathrm{S}}(\mathrm{r_{1},r_{2}})
+ \gamma_{\mu}^{(1)}\gamma_{\mu}^{(2)}
f_{k}^{\mathrm{V}}(\mathrm{r_{1},r_{2}})\right] 
\end{equation}

\noindent where
\begin{equation}
f_{k}^{\mathrm{S}}(\mathrm{r_{1},r_{2}}) = 
- \mathrm{\frac{g_{S}g_{S\Lambda}}{4\pi}}
(2k+1)\mathrm{\frac{2m_{S}}{\pi}}\mathrm{i}_{k}(\mathrm{m_{S}r_{<}})
\mathrm{k}_{k}(\mathrm{m_{S}r_{>}}) 
\end{equation}
\begin{equation}
f_{k}^{\mathrm{V}}(\mathrm{r_{1},r_{2}}) = 
\mathrm{\frac{g_{V}g_{V\Lambda}}{4\pi}}
(2k+1)\mathrm{\frac{2m_{V}}{\pi}}\mathrm{i}_{k}(\mathrm{m_{V}r_{<}})
\mathrm{k}_{k}(\mathrm{m_{V}r_{>}}) 
\end{equation}

\noindent where $\mathrm{r}_{<}$ ($\mathrm{r}_{>}$) is the smaller (larger)
of $\mathrm{r_{1}}$ and $\mathrm{r_{2}}$. 
Here $\mathrm{i}_{k}(\mathrm{mr})$ and $\mathrm{k}_{k}(\mathrm{mr})$ 
are modified spherical Bessel functions of order $k$. 

The matrix elements in Eqs.\ (\ref{eqn:A1}) and (\ref{eqn:A2})
are actually six dimensional integrals.  
Treating the $\gamma$-matrices as $2 \times 2$ block
matrices operating on the upper and lower components of the Hartree
spinors, these Dirac matrix elements, for each term in the 
interaction, are actually the sum of
four separate integrals. The scalar and vector time ($\mu = 4$) 
components of the effective interaction take the form of Eq.\ (\ref{eqn:A1});
the vector spatial ($\mu = 1,2,3$) components take the form of
Eq.\ (\ref{eqn:A2}). Thankfully, angular momentum relations allow 
one to integrate out the angular dependence \cite{ref:Ed57}. 
These integrals, for the scalar and vector time components, become
\begin{eqnarray}
& (51) = & {\displaystyle\sum_{k=0}^{\infty}} 
\langle12|f_{k}^{i}(\mathrm{r_{1},r_{2}})|34\rangle
(-1)^{\mathrm{j_{2}+j_{3}+J}} \left\{
\begin{array}{ccc}
\mathrm{J} & \mathrm{j_{2}} & \mathrm{j_{1}} \\
k & \mathrm{j_{3}} & \mathrm{j_{4}} \\
\end{array} \right\} \mathrm{\delta_{JJ'}\delta_{MM'}}
\nonumber \\ & & \times \langle (l_{1}\frac{1}{2})\mathrm{j_{1}}||
\mathrm{C_{\mathit{k}}(1)}||(l_{3}\frac{1}{2})\mathrm{j_{3}}\rangle 
\langle (l_{2}\frac{1}{2})\mathrm{j_{2}}||\mathrm{C_{\mathit{k}}(2)}||
(l_{4}\frac{1}{2})\mathrm{j_{4}}\rangle
\label{eqn:A10}
\end{eqnarray}

\noindent where $i =\mathrm{S,V}$ and (51) indicates the quantity
in Eq.\ (\ref{eqn:A1}). For the vector spatial 
components, these integrals become
\begin{eqnarray}
& (52) = & {\displaystyle\sum_{k=0}^{\infty}\sum_{\lambda}}
\langle12|f_{k}^{\mathrm{V}}(\mathrm{r_{1},r_{2}})
|34\rangle (-1)^{k+1-\lambda}
(-1)^{\mathrm{j_{2}+j_{3}+J}} \left\{
\begin{array}{ccc}
\mathrm{J} & \mathrm{j_{2}} & \mathrm{j_{1}} \\
\lambda & \mathrm{j_{3}} & \mathrm{j_{4}} \\
\end{array} \right\} \nonumber \\ & & \times \mathrm{\delta_{JJ'}\delta_{MM'}}
\langle (l_{1}\frac{1}{2})\mathrm{j_{1}}||
\chi_{\lambda}^{(k,1)}(1)||(l_{3}\frac{1}{2})\mathrm{j_{3}}\rangle
\langle (l_{2}\frac{1}{2})\mathrm{j_{2}}||\chi_{\lambda}^{(k,1)}(2)||
(l_{4}\frac{1}{2})\mathrm{j_{4}}\rangle \nonumber \\ & &
\label{eqn:A11}
\end{eqnarray}

\noindent The 6-j symbols limit the possible allowed 
values of $k$ and $\lambda$. The reduced matrix elements are
evaluated using \cite{ref:Ed57} and further limit $k$ 
and $\lambda$. Note that as the upper and lower Hartree spinors 
have different $l$ values, the reduced matrix elements in Eqs.\ 
(\ref{eqn:A10}) and (\ref{eqn:A11}) must have the corresponding, 
appropriate $l$ values.

Now consider the remaining two-dimensional radial integrals, 
where the numbers are a shorthand for all the quantum numbers
needed to uniquely specify the radial wave functions \cite{ref:Fu85},
\begin{equation}
\langle12|f_{k}^{i}(\mathrm{r_{1},r_{2}})|34\rangle = 
\int_{0}^{\infty}\!\!\int_{0}^{\infty}
\mathrm{dr_{1}dr_{2}U_{1}(r_{1})U_{2}(r_{2})}
f_{k}^{i}(\mathrm{r_{1},r_{2}})\mathrm{U_{3}(r_{1})U_{4}(r_{2})}
\end{equation}

\noindent Here $\mathrm{R(r) = U(r) / r}$ are the appropriate 
radial Dirac wave functions, in terms of $\mathrm{G}_{a}(\mathrm{r})$ 
and $\mathrm{F}_{a}(\mathrm{r})$, and again $i = \mathrm{S,V}$.

Using the Hartree spinor representation, the 
particle-hole matrix element is expressed as 
a sum of Dirac matrix elements of the types shown above \cite{ref:Fe71}
\begin{equation}
v_{\mathrm{ab;lm}}^{\mathrm{J}}=
\mathrm{\sum_{J'}(2J'+1}) \left\{
\begin{array}{ccc}
\mathrm{j_{m}} & \mathrm{j_{a}} & \mathrm{J'} \\
\mathrm{j_{b}} & \mathrm{j_{l}} & \mathrm{J} \\
\end{array} \right\}
\mathrm{\langle lbJ'|V|amJ'\rangle}
\end{equation} 

\noindent No exchange term is required, due to the 
fact that the $\Lambda$ and the nucleon are distinguishable particles
here. For example, the particle-hole matrix element for the 
vector spatial component of the effective interaction is 
\begin{eqnarray}
v\mathrm{_{32;14}^{J}(vs)} & = & \mathrm{(-1)^{j_{2}+j_{3}+J}}
\sum_{k}^{\infty}\sum_{\lambda}(-1)^{k}
\left\{\begin{array}{ccc}
\mathrm{j_{2}} & \mathrm{j_{4}} & \lambda \\
\mathrm{j_{1}} & \mathrm{j_{3}} & \mathrm{J} \\
\end{array} \right\} \mathrm{\int\int dr_{1}dr_{2}} 
\nonumber \\ & & \times \left\{ \mathrm{G_{1}(r_{1})
F_{3}(r_{1})}f_{k}\mathrm{^{V}(r_{1},r_{2})}\mathrm{G_{2}(r_{2})F_{4}(r_{2})}
\right. \nonumber \\ & & \times \langle 
(l_{\mathrm{1A}}\frac{1}{2})\mathrm{j_{1}}||\chi_{\lambda}^{(k,1)}(1)
||(l_{\mathrm{3B}}\frac{1}{2})\mathrm{j_{3}}\rangle
\langle(l_{\mathrm{2A}}\frac{1}{2})\mathrm{j_{2}}||\chi_{\lambda}^{(k,1)}(2)
||(l_{\mathrm{4B}}\frac{1}{2})\mathrm{j_{4}}\rangle \nonumber \\
& - & \mathrm{G_{1}(r_{1})F_{3}(r_{1})}f_{k}\mathrm{^{V}(r_{1},r_{2})}
\mathrm{F_{2}(r_{2})G_{4}(r_{2})}
\nonumber \\ & & \times \langle (l_{\mathrm{1A}}\frac{1}{2})\mathrm{j_{1}}||
\chi_{\lambda}^{(k,1)}(1)||(l_{\mathrm{3B}}\frac{1}{2})\mathrm{j_{3}}\rangle
\langle (l_{\mathrm{2B}}\frac{1}{2})\mathrm{j_{2}}||\chi_{\lambda}^{(k,1)}(2)
||(l_{\mathrm{4A}}\frac{1}{2})\mathrm{j_{4}}\rangle \nonumber \\ 
& - & \mathrm{F_{1}(r_{1})G_{3}(r_{1})}f_{k}\mathrm{^{V}(r_{1},r_{2})}
\mathrm{G_{2}(r_{2})F_{4}(r_{2})}  
\nonumber \\ & & \times \langle (l_{\mathrm{1B}}\frac{1}{2})\mathrm{j_{1}}||
\chi_{\lambda}^{(k,1)}(1)||(l_{\mathrm{3A}}\frac{1}{2})\mathrm{j_{3}}\rangle
\langle (l_{\mathrm{2A}}\frac{1}{2})\mathrm{j_{2}}||\chi_{\lambda}^{(k,1)}(2)
||(l_{\mathrm{4B}}\frac{1}{2})\mathrm{j_{4}}\rangle \nonumber \\ 
& + & \mathrm{F_{1}(r_{1})G_{3}(r_{1})}f_{k}\mathrm{^{V}(r_{1},r_{2})}
\mathrm{F_{2}(r_{2})G_{4}(r_{2})}  
\nonumber \\ & & \times \left. \langle 
(l_{\mathrm{1B}}\frac{1}{2})\mathrm{j_{1}}||
\chi_{\lambda}^{(k,1)}(1)||(l_{\mathrm{3A}}\frac{1}{2})\mathrm{j_{3}}\rangle
\langle(l_{\mathrm{2B}}\frac{1}{2})\mathrm{j_{2}}||\chi_{\lambda}^{(k,1)}(2)
||(l_{\mathrm{4A}}\frac{1}{2})\mathrm{j_{4}}\rangle \right\} \nonumber \\ & &
\label{eqn:ph1}
\end{eqnarray} 

\noindent Here $l_{i\mathrm{A}}$ and $l_{i\mathrm{B}}$ 
are the $l$ values corresponding to the upper and lower Hartree spinors
respectively for the $i$th wave function where $i=1 \ldotp\ldotp 4$.
Now the splitting, for a $\mathrm{s}_{1/2}$-doublet,
is just the difference between the particle-hole matrix 
elements of the two available states, or
\begin{equation}
\delta\epsilon=v_{\mathrm{n\Lambda;n\Lambda}}^{\mathrm{J=j_{1}+j_{2}}}
-v_{\mathrm{n\Lambda;n\Lambda}}^{\mathrm{J=|j_{1}-j_{2}|}}
\label{eqn:spl}
\end{equation}

\noindent The substitutions used to acquire the appropriate indices
for this case are $n=1,3$ and $\Lambda = 2,4$.
The solution to the Hartree equations yields a single-particle 
energy level for the GS, $\mathrm{E}_{\Lambda}$. As previously
mentioned, for the cases under consideration this level 
is in fact a doublet; however, Eq.\ (\ref{eqn:spl}) 
evaluates only the size of the splitting.
In order to determine the position of the doublet relative to 
$\mathrm{E}_{\Lambda}$, one needs the relation
\begin{equation}
\mathrm{\sum_{J}\left(2J+1\right)\delta\epsilon=0}
\end{equation}

We now have a framework with which to calculate the
size of the $\mathrm{s}_{1/2}$-splittings 
of the single $\Lambda$-hypernuclei of interest here and to determine
their location relative to $\mathrm{E}_{\Lambda}$. 
The problem is reduced to Slater integrals and some algebra;
the 6-j and 9-j symbols are determined using \cite{ref:Ma55,ref:Ro59}. 
The Dirac wave functions needed to solve the radial integrals are taken as 
the solutions to the Hartree equations from the previous section. 
Once all the parameters in the underlying lagrangian are
fixed, the splitting is completely determined in this approach
as there are no additional constants fit to excited state
properties \cite{ref:Fu85}. We also mention that this approach is 
applicable to excited states and multiplets for this class of 
nuclei.

To calibrate this approach, we apply it to 
ordinary nuclei. Two modifications to our framework
are required here. First, an exchange term is included because
the proton and neutron are indistinguishable particles.
As a result, the particle-hole matrix element becomes
the following \cite{ref:Fe71}
\begin{eqnarray}
v_{\mathrm{ab;lm}}^{\mathrm{J}} & = &
\mathrm{\sum_{J'}(2J'+1)} \left\{
\begin{array}{ccc}
\mathrm{j_{m}} & \mathrm{j_{a}} & \mathrm{J'} \\
\mathrm{j_{b}} & \mathrm{j_{l}} & \mathrm{J} \\
\end{array} \right\} \nonumber \\ & & \times
\left[\mathrm{\langle lbJ'|V|amJ'\rangle} - \mathrm{(-1)^{j_{a}+j_{m}+J'}}
\mathrm{\langle lbJ'|V|maJ'\rangle}\right]
\end{eqnarray} 

\noindent Second, the effective interaction is also modified, 
requiring additional couplings
to the rho and pion fields \cite{ref:Fu85}
\begin{eqnarray}
\mathrm{V(r_{12})} & = & \gamma_{4}^{(1)}\gamma_{4}^{(2)}\left[
\mathrm{\frac{-g_{S}^{2}}{4\pi}\frac{e^{-m_{S}r_{12}}}{r_{12}}
+\gamma_{\mu}^{(1)}\gamma_{\mu}^{(2)}
\frac{g_{V}^{2}}{4\pi}\frac{e^{-m_{V}r_{12}}}{r_{12}}}\right.
\nonumber \\ & + & \left. \mathrm{\gamma_{\mu}^{(1)}\gamma_{\mu}^{(2)}
\frac{\vec{\tau}^{(1)}\cdot\vec{\tau}^{(2)}}{4}
\frac{g_{\rho}^{2}}{4\pi}\frac{e^{-m_{\rho}r_{12}}}{r_{12}}}
+ \mathrm{\gamma_{5}^{(1)}\gamma_{5}^{(2)}\vec{\tau}^{(1)}\cdot\vec{\tau}^{(2)}
\frac{g_{\pi}^{2}}{4\pi}\frac{e^{-m_{\pi}r_{12}}}{r_{12}}}\right]
\nonumber \\ & & 
\end{eqnarray}

\noindent These alterations make the ordinary nuclear matter case
considerably more complicated than the case of single 
$\Lambda$-hypernuclei.

\section{Results \label{sec:5}}
\subsection{Parameter Fits}

\begin{table}
\begin{center}
\begin{tabular}{|c|c|c|c|c|} \hline
\multicolumn{4}{|c|}{Experimental Data} & M2 Calculation \\ \hline\hline
GS E/B & $^{13}_{\Lambda}\mathrm{C}$   & $-11.69 \pm 0.12$ & 
\cite{ref:Ca74} & -10.89 \\ \hline
       & $^{16}_{\Lambda}\mathrm{O}$   & $-12.50 \pm 0.35$ & 
\cite{ref:Pi91} & -12.03 \\ \hline
       & $^{28}_{\Lambda}\mathrm{Si}$  & $-16.60 \pm 0.2$ & 
\cite{ref:Ha96} & -17.37 \\ \hline
       & $^{32}_{\Lambda}\mathrm{S}$   & $-17.50 \pm 0.5$  & 
\cite{ref:Be79} & -17.95 \\ \hline
       & $^{40}_{\Lambda}\mathrm{Ca}$  & $-18.70 \pm 1.1$  & 
\cite{ref:Pi91} & -18.63 \\ \hline
       & $^{208}_{\Lambda}\mathrm{Pb}$ & $-26.5 \pm 0.5$   & 
\cite{ref:Ha96} & -27.81 \\ \hline\hline
$\mathrm{E_{SO}}$ & $^{13}_{\Lambda}\mathrm{C}$  & 
$0.15 \pm 0.09$  & \cite{ref:Ko02} & 0.150 \\ \hline\hline
$\mathrm{E_{SP}}$ & $^{13}_{\Lambda}\mathrm{C}$  & 
$10.83 \pm 0.03$ & \cite{ref:Ko02} & 8.849 \\ \hline
                                       & $^{16}_{\Lambda}\mathrm{O}$  & 
$10.6 \pm 0.1$   & \cite{ref:Ta00} & 8.314 \\ \hline
                                       & $^{40}_{\Lambda}\mathrm{Ca}$ & 
$7.70 \pm 1.0$   & \cite{ref:Be79} & 7.832 \\ \hline
\end{tabular}
\caption{The experimental data used in the 
parameter fits. This includes six GS binding energies $(\mathrm{E / B})$, 
one spin-orbit splitting of the p-states $(\mathrm{E_{SO} = |E_{1p_{1/2}} 
- E_{1p_{3/2}}|})$, and three s-p shell $\Lambda$ excitation 
energies $(\mathrm{E_{SP} = |E_{1p_{3/2}} - E_{1s_{1/2}}|})$.
The calculated values of these observables, using the M2 set, 
are also shown. These values are given in MeV.}
\label{tab:data}
\end{center}
\end{table}

\begin{table}
\begin{center}
\begin{tabular}{|c|c|c|c|c|c|} \hline\hline
 & M1 & M2 & M3-1 & M3-2 & M4 \\ \hline
$\mathrm{g_{S\Lambda} / g_{S}}$ & 0.87357 & 0.87697 & 0.87390  &  
0.87090 & 0.87697 \\ \hline
$\mathrm{g_{V\Lambda} / g_{V}}$ & 1.0     & 0.98623 & 0.97766  &  
0.98050 & 0.98623 \\ \hline
$\mathrm{g_{T\Lambda} / 4}$     &         & -0.892  & -0.891   &  
-0.877  & -0.892 \\ \hline
$\mu_{1}$                       &         &         & -0.1550  &  
0.1500  & 0.0700  \\ \hline
$\mu_{2}$                       &         &         & -0.2517  &  
0.2436  & 0.3111  \\ \hline
$\mu_{3}$                       &         &         &          &  
        & 0.0700  \\ \hline
\end{tabular}
\caption{Lists of the constants in the five parameter sets 
constructed here. Note that all the constants are natural and that
these sets represent different levels of sophistication in the
$\Lambda$-lagrangian.}
\label{tab:parfit}
\end{center}
\end{table}

The full lagrangian contains a number of free parameters. 
Those constants which lie in the nucleon and meson sectors 
are fixed by the G2 parameter set of FST \cite{ref:Fu97}, given in 
Table \ref{tab:PAR}. In the $\Lambda$ sector of the lagrangian, 
a total of six parameters remain undetermined to order $\nu=3$.
Fits are conducted at various levels 
of truncation in the underlying $\Lambda$-lagrangian to fix the relevant 
constants. The fits performed here are entirely separate from the one which 
determined the G2 parameter set; however, the framework which FST 
used to conduct their fits is identical to the one employed here.
The experimental data utilized to constrain the parameters in 
the $\Lambda$-lagrangian is listed in Table \ref{tab:data} and  
consists of three types of observables: GS
binding energies, s-p shell $\Lambda$ excitation energies, and spin-orbit
splittings of the p-states. Now we use the framework outlined
in sections \ref{sec:2} and \ref{sec:3} to calculate these same
observables for some initial guess of the parameters. The calculated and
experimental values are both substituted into the equation
\begin{equation}
\chi^{2}_{\mathrm{N}} = \sum_{i}\sum_{\mathrm{X}}\left[\frac{\mathrm{X}^{(i)}
_{\mathrm{exp}}-\mathrm{X}^{(i)}_{\mathrm{th}}}
{\mathrm{W}^{(i)}_{\mathrm{X}}\mathrm{X}^{(i)}_{\mathrm{exp}}}\right]^{2}
\label{eqn:fit1}
\end{equation}

\noindent where N is the number of data points and 
$\mathrm{W}^{(i)}_{\mathrm{X}}$ are the weights.
The parameters are varied such that the theoretical
and experimental values converge. The constants are fixed at
the values that produce a minimum in $\chi^{2}_{\mathrm{N}}$.

Our underlying $\Lambda$-lagrangian is truncated at four 
different levels and separate parameter fits are conducted 
at each. First, we consider the simplest possible case; only 
terms to order $\nu=2$ are retained in the 
$\Lambda$-lagrangian, which corresponds to
${\cal L}_{\Lambda}^{(2)}$. This $\Lambda$-lagrangian has a total
of two free parameters, $\mathrm{g_{S\Lambda}}$ and
$\mathrm{g_{V\Lambda}}$. In this case, the vector coupling is assumed
to be universal, as it is coupled to the conserved baryon current,
and the scalar coupling is fit to reproduce the binding energy
of a single  $\Lambda$ in nuclear matter, which is about -28 MeV 
\cite{ref:Ga88}. These assumptions are in keeping with the previous
work in \cite{ref:Mc02}. The parameters determined here are shown in Table 
\ref{tab:parfit} as the M1 set.
This set reproduces the GS binding energies fairly well, but
is unable to simulate either the correct spin-orbit splitting in the
p-states or the s-p shell excitation energies in light
$\Lambda$-hypernuclei.

\begin{figure}
\begin{center}
\includegraphics*[width=8.5 cm]{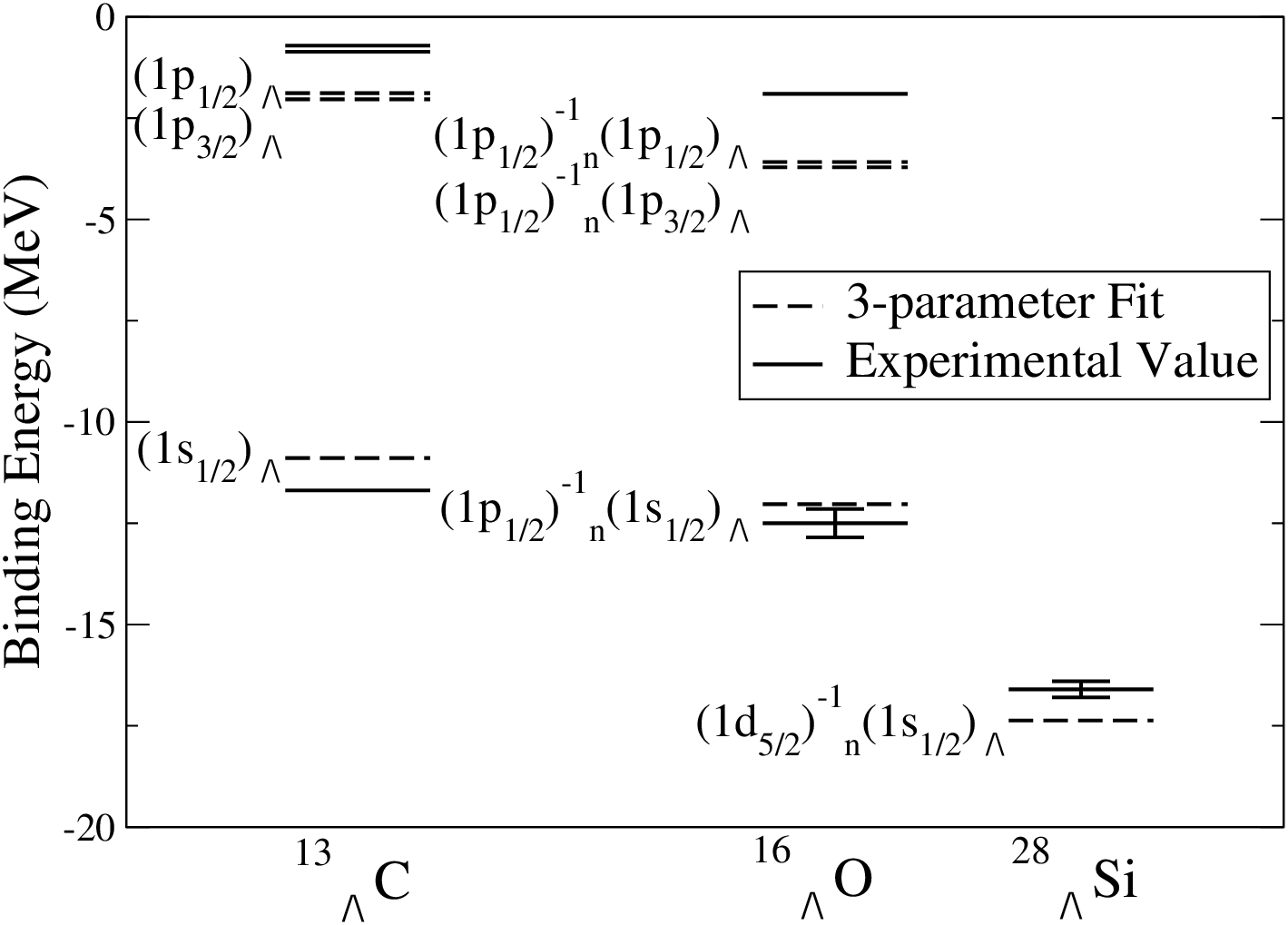}
\includegraphics*[width=8.5 cm]{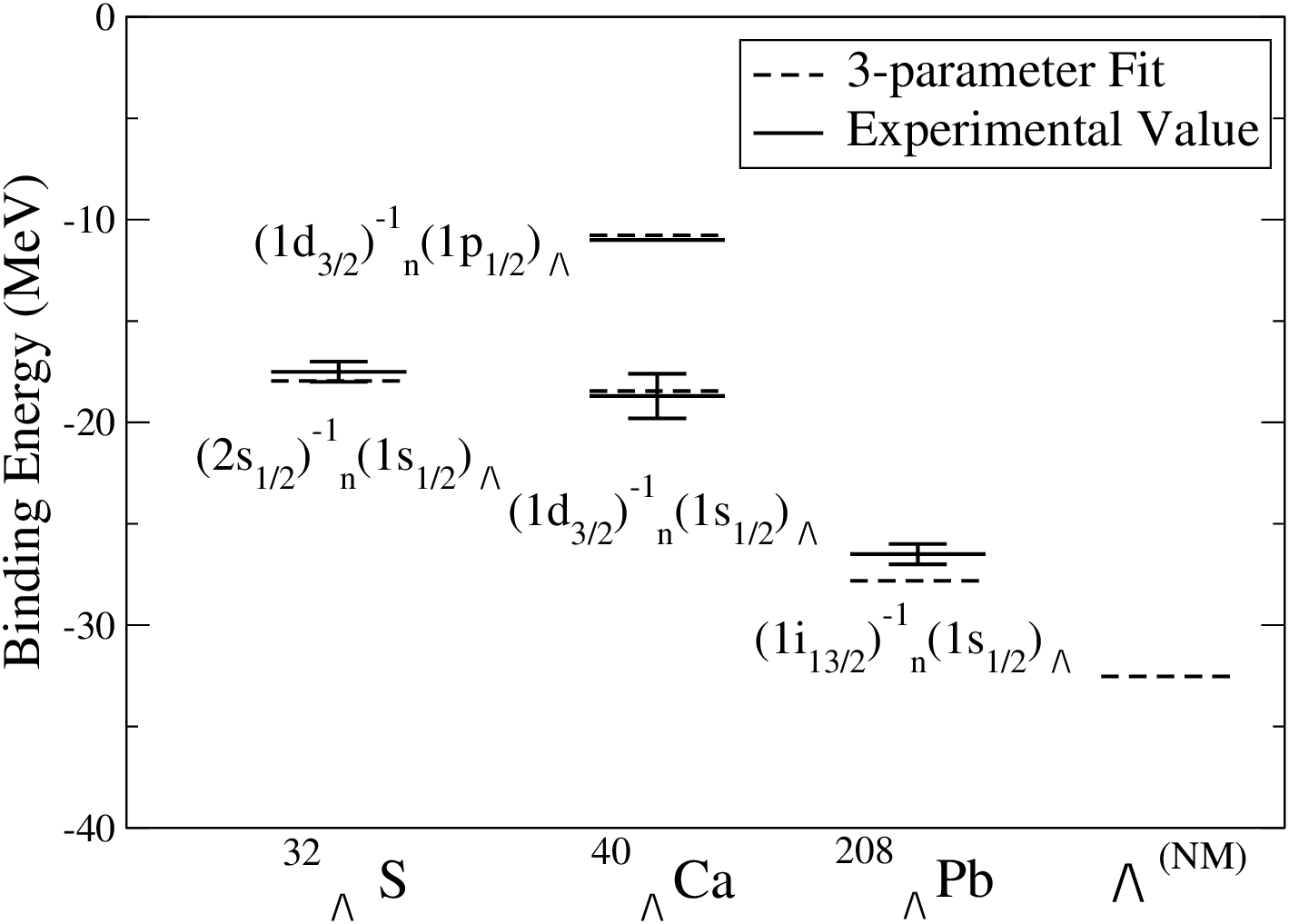}
\caption{Results of the unweighted 3-parameter fit to a series of experimental 
data. The G2 parameter set of FST is used for both the nucleon and meson
sectors \cite{ref:Fu97}. The calculated binding energy of a single $\Lambda$
in infinite nuclear matter is also shown.}
\label{fig:3parfit1}
\end{center}
\end{figure}

\begin{table}
\begin{center}
\begin{tabular}{|c|c|c|c|c|} \hline\hline
 & M2 & M3-1 & M3-2 & M4 \\ \hline
$\chi^{2}_{10}(\mathrm{UW}) \times 100$ & 0.105 & 
0.0877 &         &       \\ \hline
$\chi^{2}_{10}(\mathrm{W}) \times 10$   & 0.598 &
        & 0.515 & 0.485 \\ \hline
\end{tabular}
\caption{The $\chi^{2}$ values for both the unweighted (UW) and 
weighted (W) fits relative to the $\chi^{2}$ of the M1 set. Here 
$\chi^{2}$ is determined from Eq.\ (\ref{eqn:fit1}) using 10 pieces 
of data.}
\label{tab:chi2}
\end{center}
\end{table}

In order to obtain a better fit to the data, we increase the level
of truncation. Therefore, tensor couplings to both the vector and photon fields
are included, which correspond to the terms in 
${\cal L}_{\Lambda}^{\mathrm{(T)}}$. As a result, a third free parameter, 
$\mathrm{g_{T\Lambda}}$, is introduced.
This fit is performed using seven pieces of experimental data:
the six GS binding energies and the spin-orbit splitting
given in Table \ref{tab:data}. In this particular case, the weights in 
Eq.\ (\ref{eqn:fit1}) are all taken to be equal. The resulting 
parameters are given in Table \ref{tab:parfit} as the M2 set 
and all satisfy the assumption of naturalness. Table 
\ref{tab:data} also outlines the numerical 
results of this 3-parameter fit. The outcome of this fit is 
shown graphically in Fig.\ \ref{fig:3parfit1}. 
One can see that both the GS binding energies and the small spin-orbit
splitting in the p-states are reproduced well. The
calculated s-p shell excitation energies fail to duplicate the experimental
values for the lightest $\Lambda$-hypernuclei;
however, it is correctly given by the time one gets to
$^{40}_{\Lambda}\mathrm{Ca}$. In Fig.\ \ref{fig:3parfit1},
the value of $-32.4$ MeV is given as the calculated binding
energy of a single $\Lambda$ in nuclear matter.
This M2 parameter set will be used in the subsequent 
calculation of the $\mathrm{s}_{1/2}$-splittings.
 
\begin{figure}
\begin{center}
\includegraphics*[width=8.5 cm]{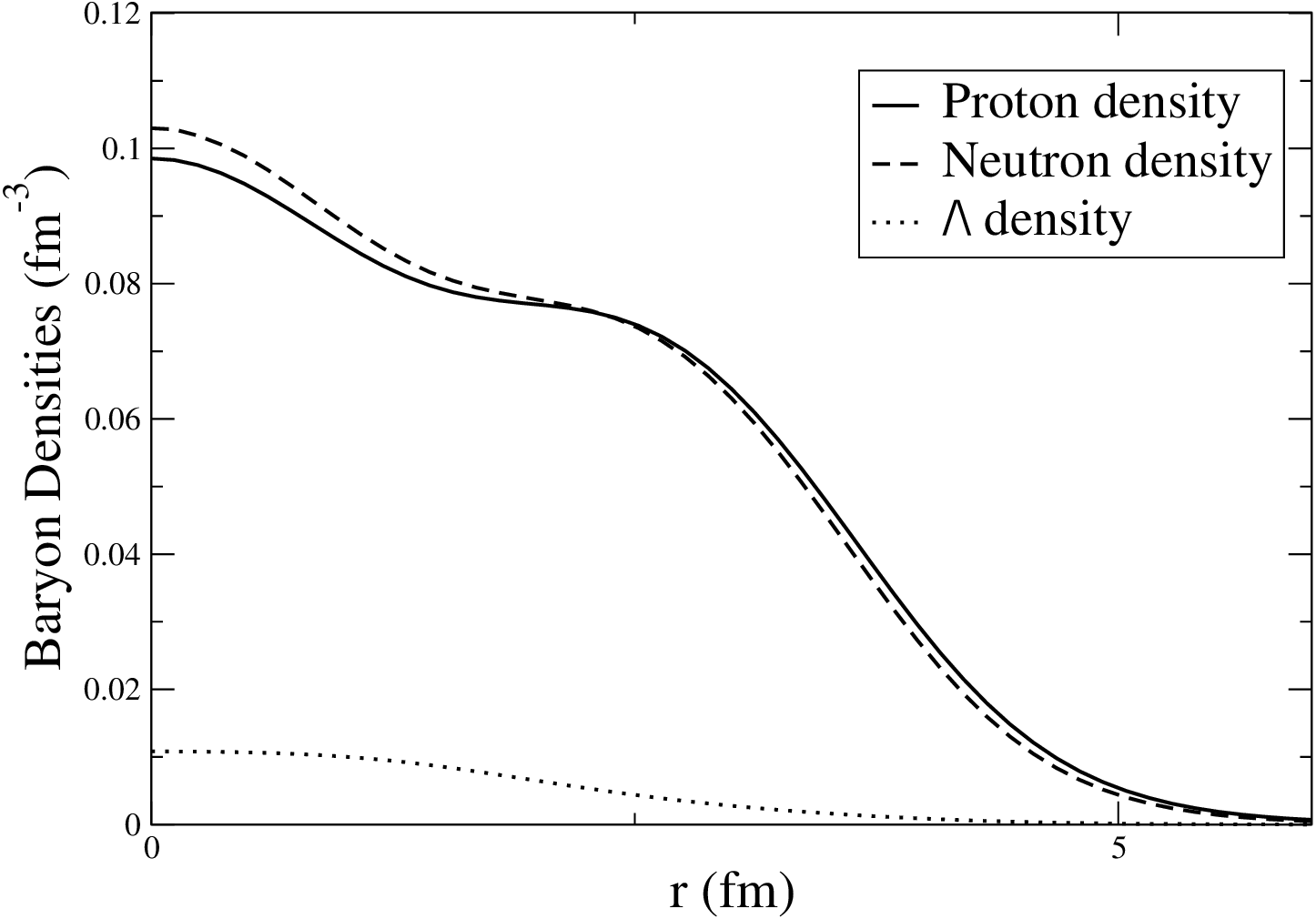}
\includegraphics*[width=8.5 cm]{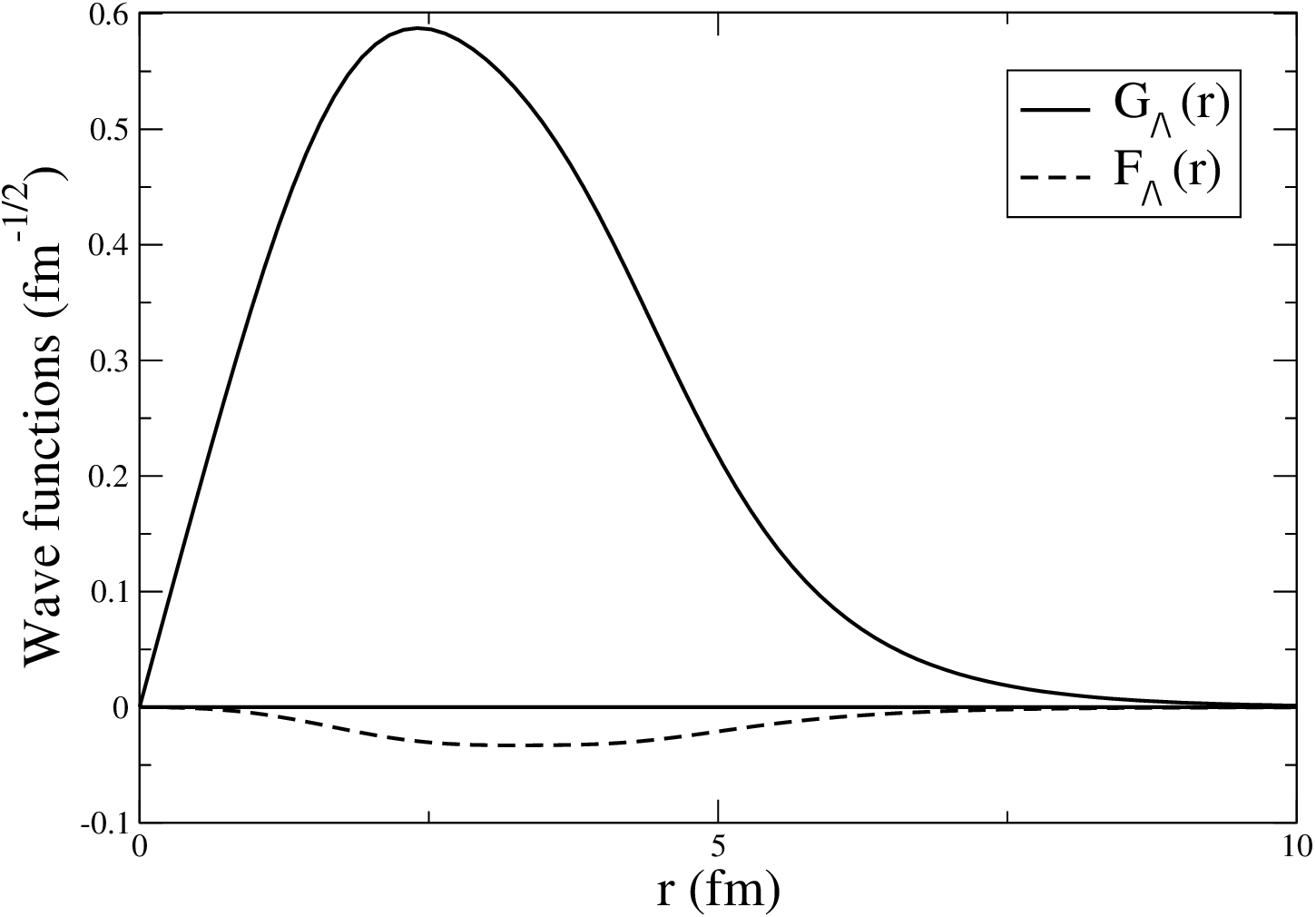}
\caption{Top: plot of the proton, neutron, and lambda densities for the GS of
$^{40}_{\Lambda}\mathrm{Ca}$. Bottom: radial wave functions of the 
$\Lambda$ in the $(1\mathrm{s}_{1/2})$ state for the GS of 
$^{40}_{\Lambda}\mathrm{Ca}$. Here the M2 parameter set was used.}
\label{fig:wfunc}
\end{center}
\end{figure}

A plot of the proton, neutron, and $\Lambda$ densities for the GS
of $^{40}_{\Lambda}\mathrm{Ca}$ calculated using this M2 set is shown 
in Fig.\ \ref{fig:wfunc}. A graph of the Hartree spinors
from the $\Lambda$ wave function, $\mathrm{G_{\Lambda}(r)}$ and
$\mathrm{F_{\Lambda}(r)}$, for the GS of $^{40}_{\Lambda}\mathrm{Ca}$
using the M2 set is also given in Fig.\ \ref{fig:wfunc}. 
Notice that the magnitude of the lower spinor is very small;
this indicates that the $\Lambda$ is essentially behaving as 
a nonrelativistic particle in the nuclear potential.

Next, the two terms nonlinear in the scalar and vector field, 
shown in ${\cal L}_{\Lambda}^{\mathrm{(N)}}$, are retained. 
This brings the number of unconstrained parameters up to five. 
For this 5-parameter fit, ten pieces of experimental
data are used; in addition to the data utilized in the 3-parameter fit,
the three s-p shell $\Lambda$ excitation energies listed in Table 
\ref{tab:data} are also included. Two versions of the 5-parameter
fit were conducted here: one unweighted and one weighted. 
In the former case, all of the weights are equal. For the latter,
the weighting scheme is as follows: $\mathrm{W}_{\mathrm{X}}^{(i)}=1.0$ 
for GS binding energies; $\mathrm{W}_{\mathrm{X}}^{(i)}=10.0$ for s-p 
shell $\Lambda$ excitation energies; and 
$\mathrm{W}_{\mathrm{X}}^{(i)}=40.0$ for the spin-orbit splitting. 
The weights were selected using the formula 
$\mathrm{W}^{(i)}_{\mathrm{X}} = f_{i}(\mathrm{\Delta E_{exp} / E_{exp}})$
where $f_{i}$ is an arbitrary factor chosen to prevent
any observable from dominating the fit \cite{ref:Ni92}. However, not enough
similar data was available to constrain the two new parameters 
individually. As a result, we initially restrict 
these parameters with the relation
\begin{equation}
\frac{\mu_{2}}{\mu_{1}} = \mathrm{\left(\frac{g_{S}\phi_{0}}
{g_{V}V_{0}}\right)_{n.m.}^{2}} = 1.624
\label{eqn:cnstrt}
\end{equation}

\noindent where n.m.\ denotes the nuclear matter values \cite{ref:Se97}. 
The results of both 5-parameter
fits are shown in Table \ref{tab:parfit}; the M3-1 and M3-2
sets denote the unweighted and weighted schemes respectively.
Again notice the parameters are all natural. However, the new parameters
are not very well determined and fail to significantly
improve the fit in either case, as can be seen from Table
\ref{tab:chi2}. Therefore, we leave the constraint
of Eq.\ (\ref{eqn:cnstrt}) intact.

Lastly, to include all possible terms in the $\Lambda$-lagrangian up to 
order $\nu = 3$, all three terms in ${\cal L}_{\Lambda}^{\mathrm{(N)}}$
are retained. Again, not enough similar data was available to individually
constrain the new parameters; therefore, we restrict these parameters
with the relation 
\begin{equation}
\mu_{1} = \mu_{3} = 0.225 \mu_{2}
\end{equation} 

\noindent and fix the remaining constants using the M2 set. 
These ratios were chosen because they tend to concentrate the 
effects of the new contributions in the surface of the nucleus, i.e.\
the additional contributions now vanish for uniform nuclear matter.
This will have a greater effect on the s-p shell excitations than
on the GSs. The weighting scheme described above was used. The resulting
parameters are listed in Table \ref{tab:parfit} as the M4 set.
Again, as seen in Table \ref{tab:chi2} the improvement in the 
overall fit is negligible. The M3-2 and M4 sets both improve the fit 
to the GSs but do worse with respect to the s-p shell excitations; the M3-1
set has the opposite effect. Also we mention that the parameter
sets M3-1, M3-2, and M4 yield very similar density distributions
to those acquired from the M2 set.

\subsection{$\mathrm{s_{1/2}}$-splittings}

In this section we discuss the calculation of the 
$\mathrm{s_{1/2}}$-splittings in $\Lambda$-hypernuclei
and the results obtained from these
calculations. Following the methodology established in section
\ref{sec:4}, one needs to evaluate $\delta\epsilon$ from
Eq.\ (\ref{eqn:spl}) to determine the size of these doublets. 
It is possible to separate $\delta\epsilon$
into contributions from each portion of the effective interaction, or
\begin{equation}
\delta\epsilon = \delta\epsilon(\mathrm{s})+\delta\epsilon(\mathrm{vt})
+\delta\epsilon(\mathrm{vs})
\end{equation}

\noindent where s, vt, and vs represent the scalar, vector time, 
and vector spatial components respectively.
As it turns out, the scalar and vector time components each cancel
in the splitting, shown by
\begin{equation}
\mathrm{\delta\epsilon(s) = \delta\epsilon(vt)} = 0
\label{eqn:zero}
\end{equation}

\noindent Therefore, the $\mathrm{s_{1/2}}$-splittings
are entirely determined from the vector spatial term in the 
effective interaction, or 
\begin{equation}
\delta\epsilon = \delta\epsilon(\mathrm{vs})
\end{equation}

\noindent This is true for any system in which either the $\Lambda$ 
or the nucleon hole has $\mathrm{j= 1/2}$. Note that this calculation
tests a different sector of the underlying lagrangian than the
mean field analysis and that, as there is no corresponding 
interpretation in the static limit $(\mathrm{M \rightarrow \infty})$,
it is here an entirely relativistic effect. Now, to determine the 
splitting we only need to evaluate the matrix element in Eq.\ 
(\ref{eqn:ph1}) for the two appropriate J values. 
The integrals are solved using the Hartree spinors, $\mathrm{G}_{a}
(\mathrm{r})$ and $\mathrm{F}_{a}(\mathrm{r})$, calculated in
the single-particle analysis. Notice that the integrals in the vector
spatial contribution mix the upper and lower components of the Hartree 
wave functions. Numerically, the integration is performed using Simpson's 
method. 
\begin{table}
\begin{center}
\begin{tabular}{|c|c|c|c|} \hline\hline
Nucleus & State & Levels & $|\delta\epsilon|$ \\ \hline
$^{12}_{\Lambda}\mathrm{B}$   & 
$\mathrm{(1p_{3/2})_{p}^{-1}(1s_{1/2})_{\Lambda}}$    &
$2^{-}_{\mathrm{GS}}$, $1^{-}$ & 426 \\ \hline
$^{16}_{\Lambda}\mathrm{N}$   & 
$\mathrm{(1p_{1/2})_{p}(1s_{1/2})_{\Lambda}}$         & 
$1^{-}_{\mathrm{GS}}$, $0^{-}$ & 472 \\ \hline
 & $\mathrm{(1p_{3/2})_{p}^{-1}(1s_{1/2})_{\Lambda}}$ & 
$2^{-}_{\mathrm{LL}}$, $1^{-}$ & 316 \\ \hline
$^{16}_{\Lambda}\mathrm{O}$   & 
$\mathrm{(1p_{1/2})_{n}(1s_{1/2})_{\Lambda}}$         & 
$1^{-}_{\mathrm{GS}}$, $0^{-}$ & 480 \\ \hline
 & $\mathrm{(1p_{1/2})_{n}(1p_{3/2})_{\Lambda}}$      &
$2^{+}_{\mathrm{LL}}$, $1^{+}$ & 125 \\ \hline
 & $\mathrm{(1p_{1/2})_{n}(1p_{1/2})_{\Lambda}}$      &
$1^{+}_{\mathrm{LL}}$, $0^{+}$ & 661 \\ \hline
$^{28}_{\Lambda}\mathrm{Si}$  & 
$\mathrm{(1d_{5/2})_{n}^{-1}(1s_{1/2})_{\Lambda}}$    &
$3^{+}_{\mathrm{GS}}$, $2^{+}$ & 293 \\ \hline
$^{32}_{\Lambda}\mathrm{S}$   & 
$\mathrm{(2s_{1/2})_{n}(1s_{1/2})_{\Lambda}}$         &
$1^{+}_{\mathrm{GS}}$, $0^{+}$ & 216 \\ \hline
$^{40}_{\Lambda}\mathrm{Ca}$  & 
$\mathrm{(1d_{3/2})_{n}^{-1}(1s_{1/2})_{\Lambda}}$    &
$2^{+}_{\mathrm{GS}}$, $1^{+}$ & 308 \\ \hline
 & $\mathrm{(1d_{3/2})_{n}^{-1}(1p_{1/2})_{\Lambda}}$ &
$2^{-}_{\mathrm{LL}}$, $1^{-}$ & 393 \\ \hline
$^{208}_{\Lambda}\mathrm{Pb}$ & 
$\mathrm{(1i_{13/2})_{n}^{-1}(1s_{1/2})_{\Lambda}}$   &
$7^{+}_{\mathrm{GS}}$, $6^{+}$ & 24  \\ \hline
\end{tabular}
\caption{$\mathrm{s_{1/2}}$-splittings, and some excited states,
are shown with their respective configurations, level orderings,
and doublet magnitudes. Here LL denotes lower level and 
$|\delta\epsilon|$ is in keV.}
\label{tab:split}
\end{center}
\end{table}

\begin{figure}
\begin{center}
\includegraphics*[width=8.5 cm]{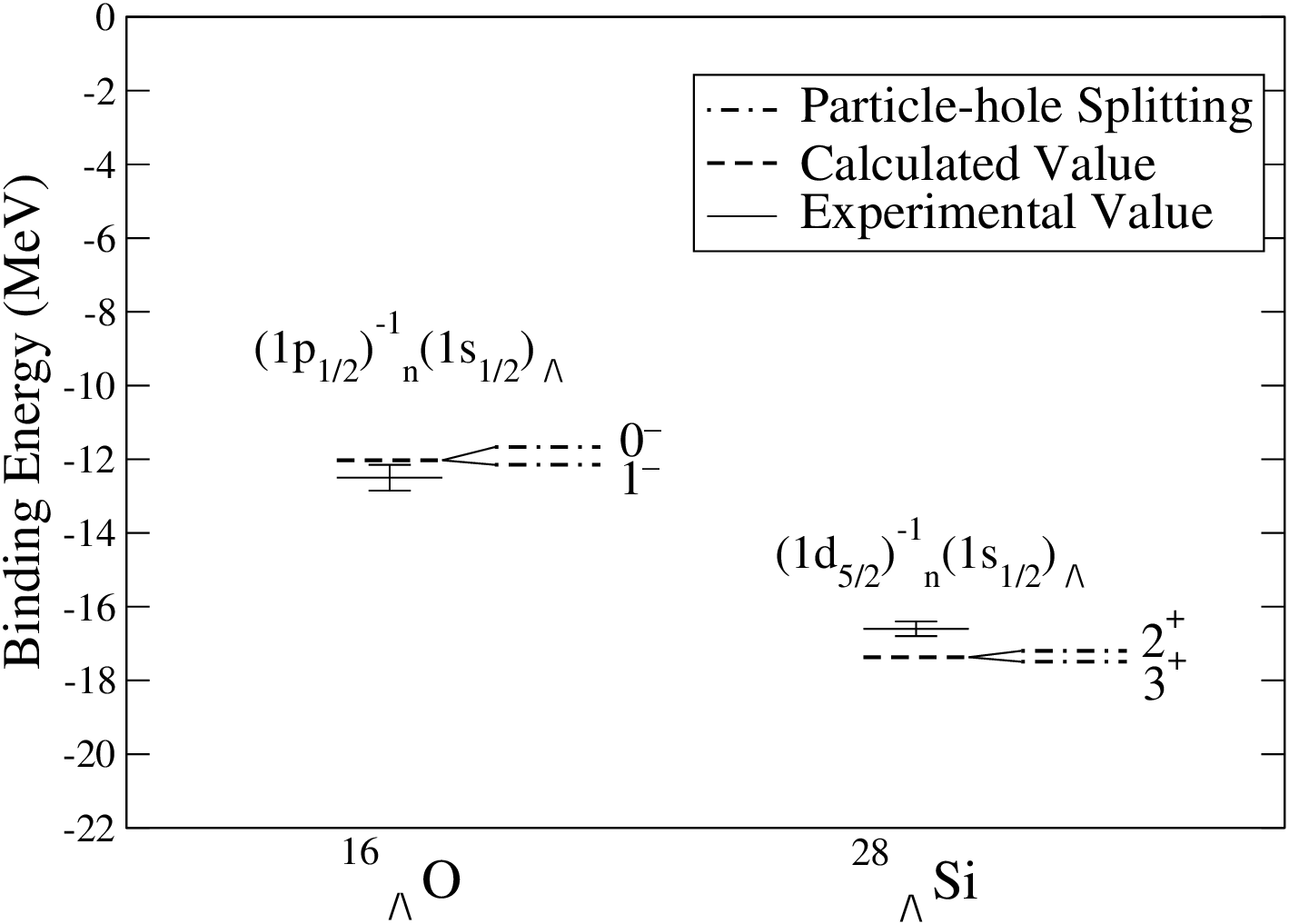}
\includegraphics*[width=8.5 cm]{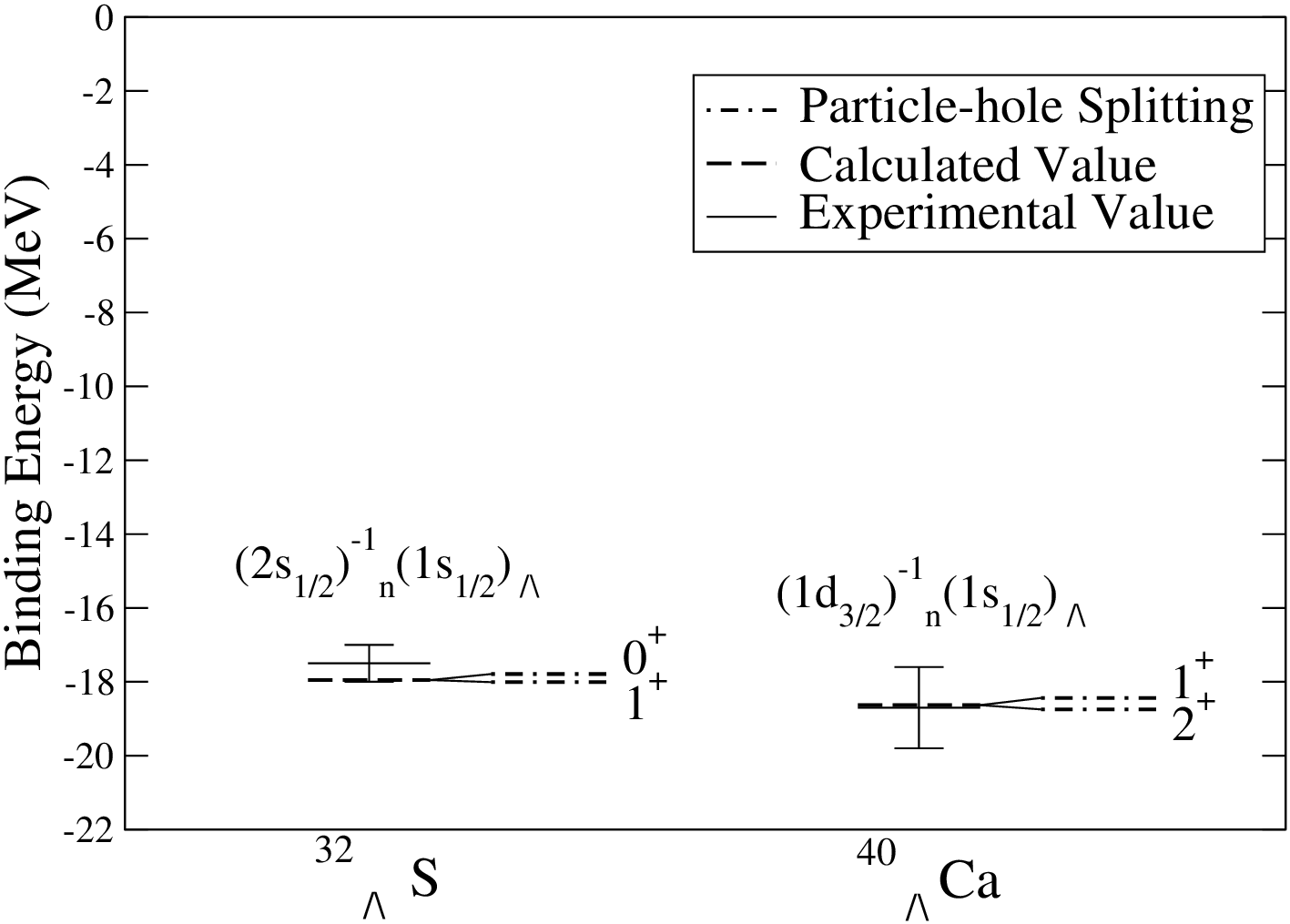}
\caption{Graph of GS particle-hole splittings and their respective 
level orderings for $^{16}_{\Lambda}\mathrm{O}$ and 
$^{28}_{\Lambda}\mathrm{Si}$ on the top and $^{32}_{\Lambda}\mathrm{S}$ and 
$^{40}_{\Lambda}\mathrm{Ca}$ on the bottom. The single-particle calculations 
were conducted using the M2 parameter set and are plotted 
alongside the experimental values \cite{ref:Pi91,ref:Be79,ref:Ha96}.
Notice that the splittings lie within the experimental error bars
in all four cases.}
\label{fig:split2}
\end{center}
\end{figure}

\begin{figure}
\begin{center}
\includegraphics*[width=8.5 cm]{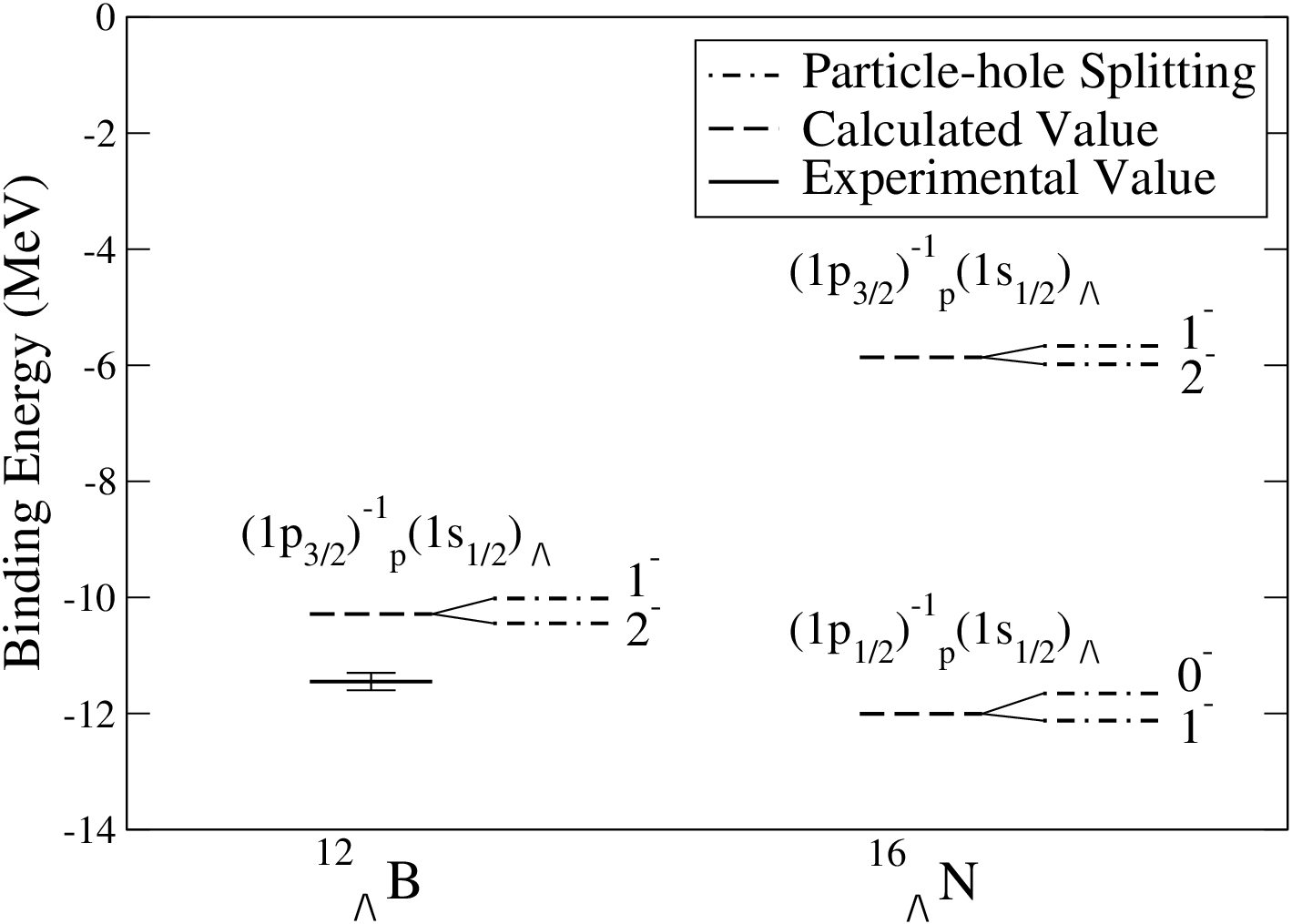}
\includegraphics*[width=8.5 cm]{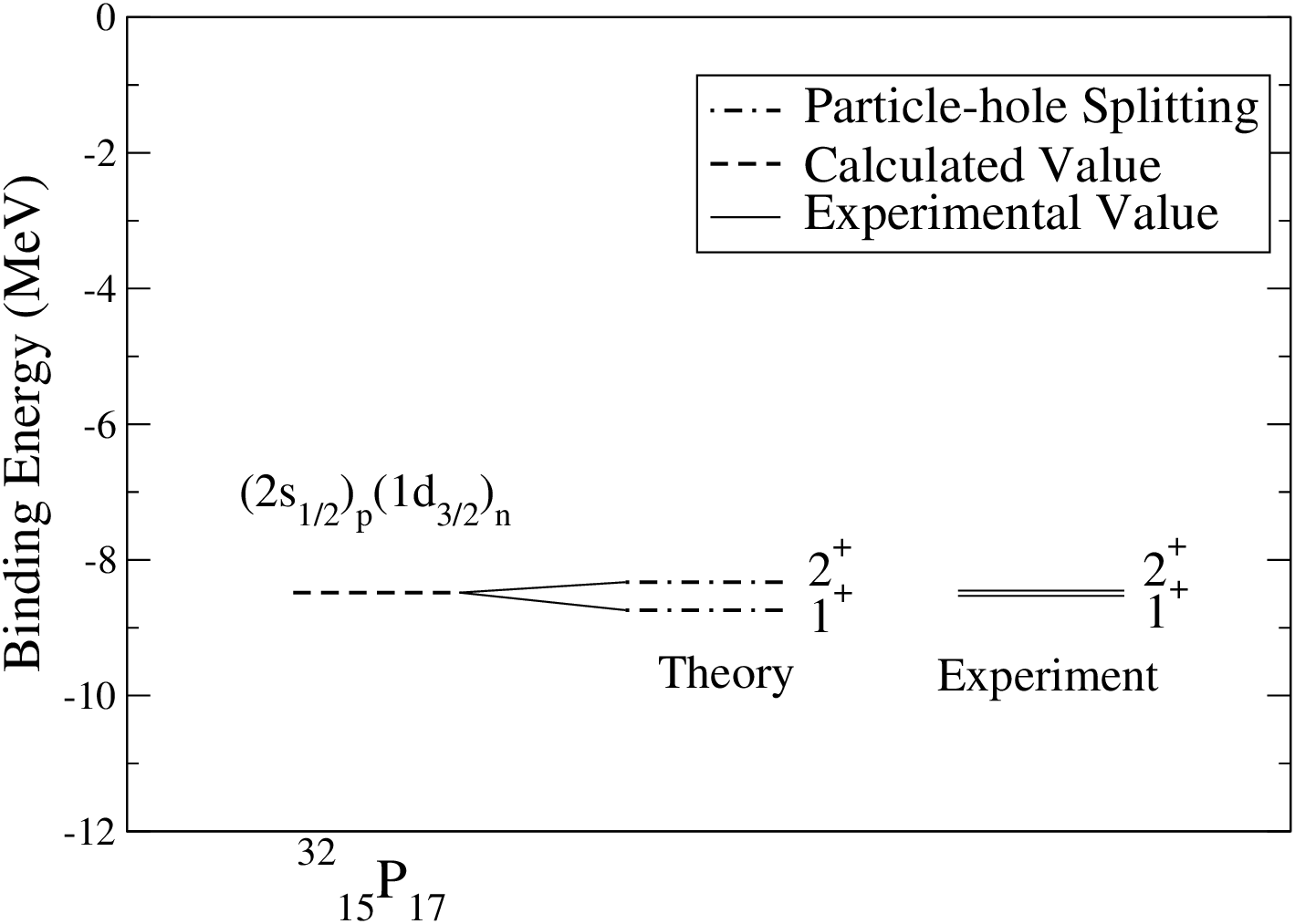}
\caption{Top: graph of particle-hole splittings for 
$^{12}_{\Lambda}\mathrm{B}$ and $^{16}_{\Lambda}\mathrm{N}$ and their 
respective level orderings. In addition to the GSs, the first calculated 
excited state in $^{16}_{\Lambda}\mathrm{N}$ is also included. These 
calculations were conducted using the M2 parameter set. The experimental 
value for the GS of $^{12}_{\Lambda}\mathrm{B}$ is taken from \cite{ref:Ju73}.
Bottom: particle-hole splitting for the GS of $^{32}_{15}\mathrm{P}_{17}$.
The level orderings and splittings are shown for both theory and 
experiment. Here the G2 parameter set of FST was used \cite{ref:Fu97}.}
\label{fig:phossplit}
\end{center}
\end{figure}

The results of this analysis are contained
in Table \ref{tab:split}. The splittings with a neutron hole listed
in Table \ref{tab:split} all correspond to single-particle
levels which were used in the fits of the preceding discussion, 
as shown in Fig.\ \ref{fig:3parfit1}. 
The $\mathrm{s_{1/2}}$-splittings for $^{16}_{\Lambda}\mathrm{O}$,
$^{28}_{\Lambda}\mathrm{Si}$, $^{32}_{\Lambda}\mathrm{S}$, and
$^{40}_{\Lambda}\mathrm{Ca}$ are plotted in Fig.\ \ref{fig:split2}; 
notice that these four splittings are all within the experimental error 
bars and that the appropriate level orderings are shown. 
It should be mentioned that the three excited states with neutron holes
shown in Table \ref{tab:split} will overlap with other states of the 
same J value. Therefore in these cases one must diagonalize the
hamiltonian to determine the correct splitting and level ordering.
The remaining doublets in
Table \ref{tab:split}, those with proton holes, are for predicted $\Lambda$
single-particle levels. These three are shown in Fig.\ \ref{fig:phossplit}; 
here, in addition to the GS splittings for both $^{12}_{\Lambda}\mathrm{B}$
and $^{16}_{\Lambda}\mathrm{N}$, the doublet for the first calculated
excited state in $^{16}_{\Lambda}\mathrm{N}$ is also given. 
These splittings will be measured in an upcoming experiment
using the reaction $\mathrm{(e,e'K^{+})}$ with much greater
resolution than the $\mathrm{(\pi^{+},K^{+})}$ reactions 
\cite{ref:Ga94,ref:Ur01}. As the 
effective interaction used here is isoscalar, there is no distinction
in this approach between proton and neutron holes. This is 
apparent when comparing the GSs of $^{16}_{\Lambda}\mathrm{N}$ and
$^{16}_{\Lambda}\mathrm{O}$; the slight difference in their splittings,
which is only about 10 keV, arises from Coulomb effects. 
Also note that the splittings for configurations with
the holes in the same shell are larger for the smaller j value. For example, 
the doublet for the GS of $^{12}_{\Lambda}\mathrm{B}$, in the
$\mathrm{(1p_{3/2})_{p}^{-1}(1s_{1/2})_{\Lambda}}$ configuration,
is smaller than that of the GS of $^{16}_{\Lambda}\mathrm{N}$,
in the $\mathrm{(1p_{1/2})_{p}^{-1}(1s_{1/2})_{\Lambda}}$
state. The level orderings for each calculated doublet are also 
given in Table \ref{tab:split}. Notice that for all of the cases 
considered here, the state with the higher J value is the GS or, 
in the case of excited states, the lower level.

Recent gamma-ray spectroscopy experiments \cite{ref:Ta03} 
(and the experimental error bars on the GS binding energy 
of $^{12}_{\Lambda}\mathrm{B}$) suggest that these particle-hole 
splittings are in fact much smaller. In addition, the measured GS
spins of $^{12}_{\Lambda}\mathrm{B}$ and  $^{16}_{\Lambda}\mathrm{O}$
are 1 and 0 respectively \cite{ref:Ju73,ref:Mi04}, 
whereas the values predicted here is 0 and 1 respectively. As the tensor 
coupling was important in the spin-orbit splittings, it may play an 
important role in the case of the $\mathrm{s}_{1/2}$-splittings. Higher 
order terms in the effective interaction, especially those involving the 
tensor coupling to the $\Lambda$, may be required to obtain a quantitative 
description of the small $\mathrm{s}_{1/2}$-doublet splitting and the 
correct level ordering. This is left for future work.  

The present analysis was also extended to the case of ordinary nuclei.
The necessary modifications to the theory were discussed 
in section \ref{sec:4}. We apply this approach to the case of 
$^{32}_{15}\mathrm{P}_{17}$ in the $\mathrm{(2s_{1/2})_{p}(1d_{3/2})_{n}}$
state. As noted before, this calculation will require
direct and exchange contributions from the scalar, vector, rho,
and pion terms in the effective interaction. Fortunately, the statement of 
Eq.\ (\ref{eqn:zero}) holds here for the direct term 
and can be extended to include the direct rho time 
component as well. The result of our calculation is 413 keV;
the observed value is 77 keV \cite{ref:Wa71}. This is shown graphically in
Fig.\ \ref{fig:phossplit};
notice that the correct magnitude and level ordering is obtained. However, 
it should be noted that this calculation is considerably more
complicated than the $\Lambda$-N case.

In {\it summary}, we have successfully extended the hadronic 
effective field theory developed by FST to the region 
of the strangeness sector corresponding to single $\Lambda$-hypernuclei.
This framework has the intrinsic strength of directly incorporating
the following: special relativity, quantum mechanics, 
the underlying symmetry structure of QCD, and the nonlinear
realization of spontaneously broken chiral symmetry. Furthermore,
DFT provides a theoretical justification for this approach. This
lagrangian can be used for predictive purposes once all the 
free parameters are determined. As a result, it was of interest
to make a minimalist extension of this methodology in which
a single, isoscalar $\Lambda$ is added to the theory.
An appropriate $\Lambda$-lagrangian was constructed as an additional 
contribution to the full interacting lagrangian of FST. This 
system was solved using the Kohn-Sham analysis. Parameter fits
were conducted at various levels of sophistication in the 
$\Lambda$-lagrangian while maintaining the full FST lagrangian
with their G2 parameter set. The 3-parameter fit reproduces the
GS binding energies and small spin-orbit splittings well, but fails
to simulate fully the s-p shell excitations in the lightest
hypernuclei, although by $^{40}_{\Lambda}\mathrm{Ca}$ 
the correct excitation energy is obtained. The inclusion of 
additional parameters does not significantly
improve the quality of the fit.

Many of the GSs used in the fits were
actually particle-hole states; as a result, it was of interest to 
calculate their splittings. A methodology for examining these 
splittings was developed using Dirac two-body matrix elements 
of an effective interaction. This effective interaction followed 
directly from the underlying lagrangian and to lowest order 
corresponded to simple scalar and vector exchange.$^{\ref{foot:1}}$
Note that this lagrangian was designed to calculate other phenomena and 
there is nothing contained in it that guarantees the 
production of small particle-hole splittings. The primary
conclusion from the present analysis is that all of the results 
obtained for the $\mathrm{s}_{1/2}$-doublet splittings
used in the fitting procedure
lie within the current experimental error bars. As a 
partial calibration, a calculation of the GS particle-hole
splitting in $^{32}_{15}\mathrm{P}_{17}$, a much more
complicated case, achieved the correct
level ordering and doublet magnitude. Using this 
approach predictions were made for nuclei that will be 
measured in an upcoming $\mathrm{(e,e'K^{+})}$ experiment
at Jefferson Lab \cite{ref:Ga94,ref:Ur01}.$^{\ref{foot:1}}$

I would like to thank the following: Dr. J. D. Walecka for his support and
advice; Dr. M. Huertas for the early use of a program he wrote to solve 
the Hartree equations \cite{ref:He03a} and for his help in its modification; 
and Dr. B. D. Serot for his careful reading of the manuscript and his 
helpful comments. This work was supported in part by DOE grant 
DE-FG02-97ER41023. 

\appendix
\section{}
\label{sec:app}

In this appendix, we discuss the selection of the terms in our
$\Lambda$-lagrangian to order $\nu = 3$. It is straightforward to see 
which terms are retained to order $\nu = 2$, with the exception of
the four fermion terms. Therefore, the following is a list of all remaining
possible combinations of the fields to order $\nu = 3$, consistent
with this approach, and a short discussion
of each. 

\begin{itemize}

\item Four fermion terms in the nuclear case, 
such as $\mathrm{\bar{N}N\bar{N}N}$,
are eliminated by substituting the meson equations 
of motion into the lagrangian. Under normal circumstances
this is not feasible; however, this is allowed when the system is already in
equilibrium. Here we want to extend the framework of FST
to single $\Lambda$-hypernuclei with no additional mesons.
In this case, either $\mathrm{\bar{N}N}\bar{\Lambda}\Lambda$ or 
$\bar{\Lambda}\Lambda\bar{\Lambda}\Lambda$
can be eliminated using this method, but not both simultaneously. 
Fortunately, the second term involves self-fields of the $\Lambda$
and consequently, can be discarded. This scheme also applies 
to terms with more than four fermion fields.

\item The term $\bar{\Lambda}\sigma_{\mu\nu}\mathrm{V_{\mu\nu}}\Lambda$
is consistent with this framework.

\item The terms $\bar{\Lambda}\Lambda\phi^{2}$ and 
$\bar{\Lambda}\Lambda\mathrm{V_{\mu}}^{2}$ are consistent with this 
framework. In the nucleon sector, terms of this variety were regrouped 
using meson field redefinitions. Here the terms have different constants 
than in the nucleon case; therefore, these terms cannot simply be 
regrouped, unless additional mesons are included.

\item The term $\bar{\Lambda}\gamma_{\mu}\Lambda\phi\mathrm{V_{\mu}}$
is also retained. In the nuclear case, it was eliminated via the Dirac 
equation, but this is not possible here.

\item Next, the following term is consistent with this methodology,
but can be rewritten as 
\begin{equation}
\mathrm{\bar{\Lambda}\gamma_{\mu}\Lambda\frac{\partial\phi}{\partial x_{\mu}}}
= \mathrm{\frac{\partial}{\partial x_{\mu}}\left(\bar{\Lambda}\gamma_{\mu}
\Lambda\phi\right)} 
- \mathrm{\left[\frac{\partial}{\partial x_{\mu}}\left(\bar{\Lambda}
\gamma_{\mu}\Lambda\right)\right]\phi}
\end{equation}
The second term is a total derivative, which does not change
the action, and the third term is a four derivative of a 
conserved current, which is zero. Therefore this term can be neglected.

\item Consider the following three terms:
$\bar{\Lambda}\gamma_{\mu}\phi
\mathrm{\frac{\partial}{\partial x_{\mu}}}\Lambda$, 
$\bar{\Lambda}\gamma_{\mu}\gamma_{\nu}\mathrm{V_{\mu}
\frac{\partial}{\partial x_{\nu}}}\Lambda$, and
$\bar{\Lambda}\gamma_{\mu}\gamma_{\nu}\mathrm{
\frac{\partial}{\partial x_{\mu}}\frac{\partial}{\partial x_{\nu}}} 
\Lambda$.
The Dirac equation for the $\Lambda$ can be substituted into 
each of these to convert them into a type of term already
considered.

\item Lastly, all of the contributions with $\mathrm{A_{\mu}}$ 
are absorbed into other terms in
the same manner as like terms with $\mathrm{V_{\mu}}$. However,
the terms $\bar{\Lambda}\gamma_{\mu}\Lambda\mathrm{A_{\mu}}$ and
$\bar{\Lambda}\Lambda\mathrm{A_{\mu}^{2}}$ can be discarded
as $\mathrm{Q=0}$ for the $\Lambda$. Therefore, the only remaining 
electromagnetic term
is $\bar{\Lambda}\sigma_{\mu\nu}\mathrm{F_{\mu\nu}}\Lambda$.

\end{itemize}

\noindent Note that the parameters have yet to be
determined. When the terms are regrouped, the free parameters
can be redefined to suit our purposes.

\end{document}